\documentclass[aps,prx,preprint,onecolumn,citeautoscript,footinbib,
eqsecnum]{revtex4-1}  
\usepackage{amsmath,amssymb,bm,bbm} 
\usepackage{graphicx}  
\usepackage{color} 
\usepackage[dvipsnames]{xcolor}
\usepackage[papersize={8.5in,11in}]{geometry}
\usepackage[colorlinks=true]{hyperref}  
\usepackage{ulem}
\usepackage[mathscr]{euscript}
\usepackage[section]{placeins}
\usepackage{comment}
\hypersetup{
    bookmarks=true,         % show bookmarks bar?
    unicode=false,          % non-Latin characters 
    pdftoolbar=true,        % show Acrobat
    pdfmenubar=true,        % show Acrobat 
    pdffitwindow=false,     % window fit to page when opened
    pdfstartview={FitH},    % fits the width of the page to the window
    pdfsubject={},   % subject of the document
    pdfcreator={},   % creator of the document
    pdfproducer={}, % producer of the document
    pdfkeywords={} {} {}, % list of keywords
    pdfnewwindow=true,      % links in new window
    colorlinks=true,       % false: boxed links; true: colored links
    linkcolor=magenta, %red,          % color of internal links (change box color with linkbordercolor)
    citecolor=blue,        % color of links to bibliography
    filecolor=magenta,      % color of file links
    urlcolor=blue           % color of external links
} 

\usepackage{amsfonts}
\usepackage{upgreek}
\usepackage{slashed}
\usepackage{latexsym}

\newcommand{\beq}{\begin{equation}}
\newcommand{\eeq}{\end{equation}}
\def\bea{\begin{eqnarray}}
\def\eea{\end{eqnarray}}

\usepackage{braket}
\usepackage{comment}
\usepackage{slashed}

\renewcommand{\vec}[1]{\boldsymbol{#1}}
\DeclareMathOperator{\sign}{sign}

\begin{document}

\preprint{\href{https://arxiv.org/abs/2006.01140}{arXiv:2006.01140}}

\title{Deconfined criticality and ghost Fermi surfaces\\ at the onset of antiferromagnetism in a metal}

\author{Ya-Hui Zhang} 
\affiliation{Department of Physics, Harvard University, Cambridge, MA 02138, USA}
\author{Subir Sachdev}
\affiliation{Department of Physics, Harvard University, Cambridge, MA 02138, USA}

\date{\today}

\begin{abstract}
We propose a general theoretical framework, using two layers of ancilla qubits, for deconfined criticality  between a Fermi liquid with a large Fermi surface, and a pseudogap metal with a small Fermi surface of electron-like quasiparticles.  The pseudogap metal can be a magnetically ordered metal, or a fractionalized Fermi liquid (FL*) without magnetic order. A critical `ghost' Fermi surface emerges (alongside the large electron Fermi surface) at the transition, with the ghost fermions carrying neither spin nor charge, but minimally coupled to $(U(1) \times U(1))/Z_2$ or $(SU(2) \times U(1))/Z_2$ gauge fields. The $(U(1) \times U(1))/Z_2$ case describes simultaneous Kondo breakdown and onset of magnetic order: the two gauge fields induce nearly equal attractive and repulsive interactions between ghost Fermi surface excitations, and this competition controls the quantum criticality.
Away from the transition on the pseudogap side, the ghost Fermi surface
absorbs part of the large electron Fermi surface, and leads to a jump in the Hall co-efficient.
 We also find an example of an ``unnecessary quantum critical point'' between a metal with spin density order, and a metal with local moment magnetic order. The ghost fermions contribute an enhanced specific heat near the transition, and could also be detected in other thermal probes. We relate our results to the phases of correlated electron compounds.
\end{abstract}

\maketitle

\tableofcontents

\section{Introduction} 
\label{sec:intro}

The study of the quantum phase transition involving the onset of antiferromagnetic order in metals is a central topic in modern quantum condensed matter theory. There are applications to numerous materials, including the $f$-electron `heavy fermion' compounds \cite{StewartRMP,Coleman05,WolfleRMP,Si08,SSV05,SiRMP}, the cuprates \cite{CPLT18,ArmitageRMP,ShenRMP}, and the `115 compounds' \cite{Yuan115,Park115}. The standard theory involves a Landau-Ginzburg-Wilson
approach, obtaining an effective action for the antiferromagnetic order parameter damped by the low energy Fermi surface excitations \cite{WolfleRMP,hertz,millis}. However, a number of experiments, especially in quasi-two dimensional compounds, do not appear to be 
compatible with this approach \cite{Steglich1,Coleman2000,Steglich2000,CPLT18}.

In this paper, we shall present a `deconfined critical theory' \cite{senthil1,senthil2}, involving fractionalized excitations and gauge fields at a critical point, flanked by phases with only conventional excitations. 
Our results here go beyond our previous work \cite{Yahui1}, which only considered transitions between a phase with fractionalized excitations, to one without; in the present paper, neither phase will. have fractionalization. For the case of the metallic antiferromagnetic critical point, we realize here a theory with simultaneous Kondo breakdown and onset of magnetic order. Such a scenario has been discussed earlier \cite{SSV05}, but no specific critical theory was proposed; reviews of related ideas are in Refs.~\onlinecite{ColemanSi01,YRZ06,TsvelikRice19}.
We note that a deconfined critical theory for the onset of spin glass order in a metal was obtained recently \cite{DQCPSG} in a model with all-to-all random couplings. There will be no randomness in the models we consider here, and only short-range couplings.

\begin{figure}
\includegraphics[width=5in]{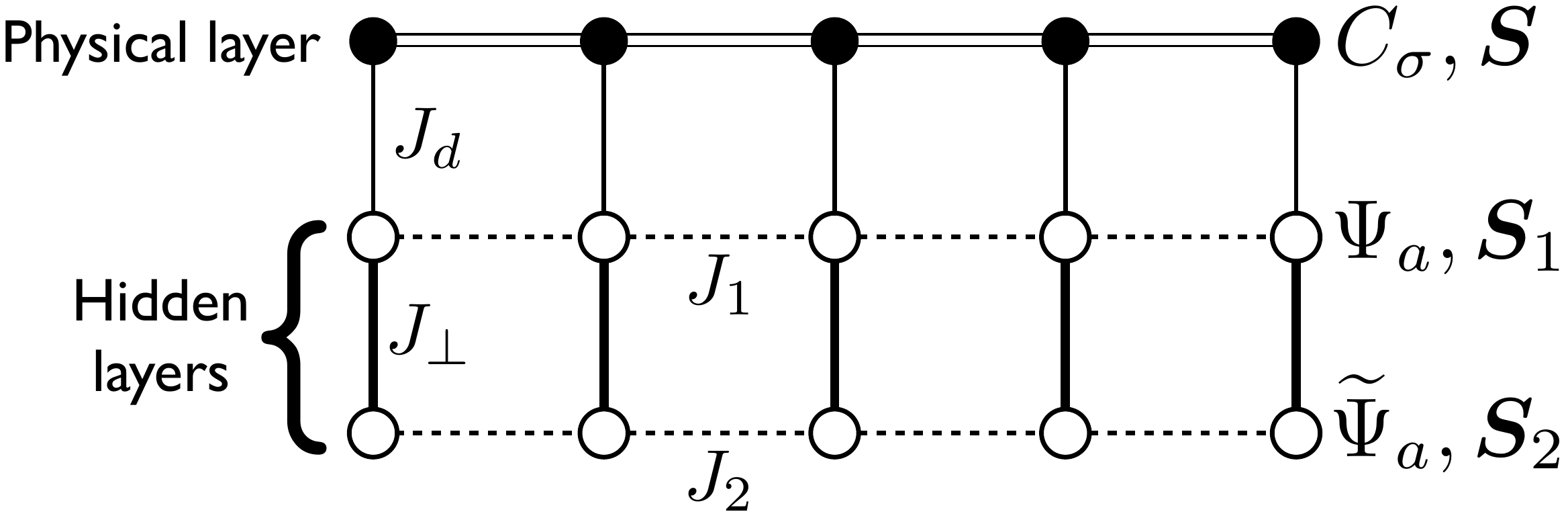}
\caption{The top layer is the physical layer coupled to two `hidden' layers of ancilla qubits (spin-1/2 spins) ${\bm S}_{i;1}$ and ${\bm S}_{i;2}$. The physical layer is taken to be either a single band model of electrons $C_{i;\sigma}$, or a Kondo lattice of electrons $C_{i ;\sigma}$ coupled to a separate set of spins ${\bm S}$.
The lattice sites are labeled by $i$, and can form any $d$-dimensional lattice (only one dimension is displayed above), and $\sigma = \uparrow,\downarrow$ is a physical spin index.
We develop a theory with the exchange interactions finite, and take the limit $J_\perp \rightarrow \infty$ of infinite antiferromagnetic exchange between the hidden layers at the end. This leads to a $(SU(2)_S \times SU(2)_1 \times SU(2)_2)/Z_2$ gauge theory, with all gauge symmetries acting on the hidden layers. The hidden layers are described by `ghost fermions' $\Psi_{i;a}$,$\widetilde{\Psi}_{i;a}$ which carry neither spin nor charge; $a=+,-$ is a $SU(2)_S$ gauge index.}
\label{fig:layers}
\end{figure}

Our theory relies on the recently introduced \cite{Yahui1} `ancilla qubit' approach
to correlated electron systems. In addition to the `physical layer' corresponding to the lattice Hamiltonian of the system of interest, we introduce two `hidden layers' of ancilla qubits (see Fig.~\ref{fig:layers}). 
By entangling the physical degrees of freedom with the ancilla, and then projecting out a trivial product state of the ancilla qubits, we are able to access a rich variety of quantum phases and critical points for the physical layer. In this manner we obtain here a deconfined theory of the onset for the antiferromagnetic order in a metal with ({\it i\/}) additional `ghost' Fermi surfaces at the critical point of fermions that carry neither spin nor charge; ({\it ii\/}) a jump in the size of the Fermi surfaces with the electron-like quasiparticles which carry both spin and charge, and a correspondingly discontinuous Hall effect; ({\it iii\/}) an enhancement of the linear in temperature specific heat at the critical point.

As we will review in Section~\ref{sec:ancilla}, the ancilla approach leads to a parent $(SU(2)_S \times SU(2)_1 \times SU(2)_2)/Z_2$ gauge theory. 
(We note that the subscript on the gauge group is merely an identifying label, and not intended to indicate the level of a Chern-Simons term; our theories here preserve time-reversal, and there are no Chern-Simons terms.)
The Higgs/confining phases, and intervening critical points or phases, of this gauge theory lead to many interesting phase diagrams of correlated electron systems. We emphasize that, unlike previous gauge theories in the literature, the electron operator in the physical layer is not fractionalized, and remains gauge neutral. All the gauge symmetries act only on the hidden layers, but we are also able to reproduce earlier results in which gauge charges resided in the physical layer. A crucial advantage of the ancilla approach is that it is far easier to keep track of Fermi surfaces, and the constraints arising from generalizations of the Luttinger constraint of Fermi liquid theory \cite{SVS04,Paramekanti_2004}.  In particular, this approach
led to \cite{Yahui1} the first self-contained description of the transformation from a fractionalized Fermi liquid (FL*) with a small Fermi surface, to a regular Fermi liquid (FL) with a large Fermi surface in a single band model, with the correct Fermi surface volumes at mean-field level in both phases. (The FL* phase has no symmetry breaking, and the violation of the conventional Luttinger constraint by its small Fermi surface is accounted for by the presence of fractionalized excitations and bulk topological order \cite{SSV03,SVS04,Paramekanti_2004}.)
The critical theory (labeled DQCP1 below) had a Fermi surface of ghost fermions and a Higgs field both carrying fundamental $(SU(2)_S \times U(1)_1)/Z_2$ gauge charges. We  note the recent work of Refs.~\onlinecite{ZouDeb1,ZouDeb2} which described  metal-insulator transition or metal-metal transition  using $U(2)$ or $U(1)\times U(1)$ gauge theories, but not using the ancilla method.

\begin{figure}
    \centering
    \includegraphics[width=0.35\textwidth]{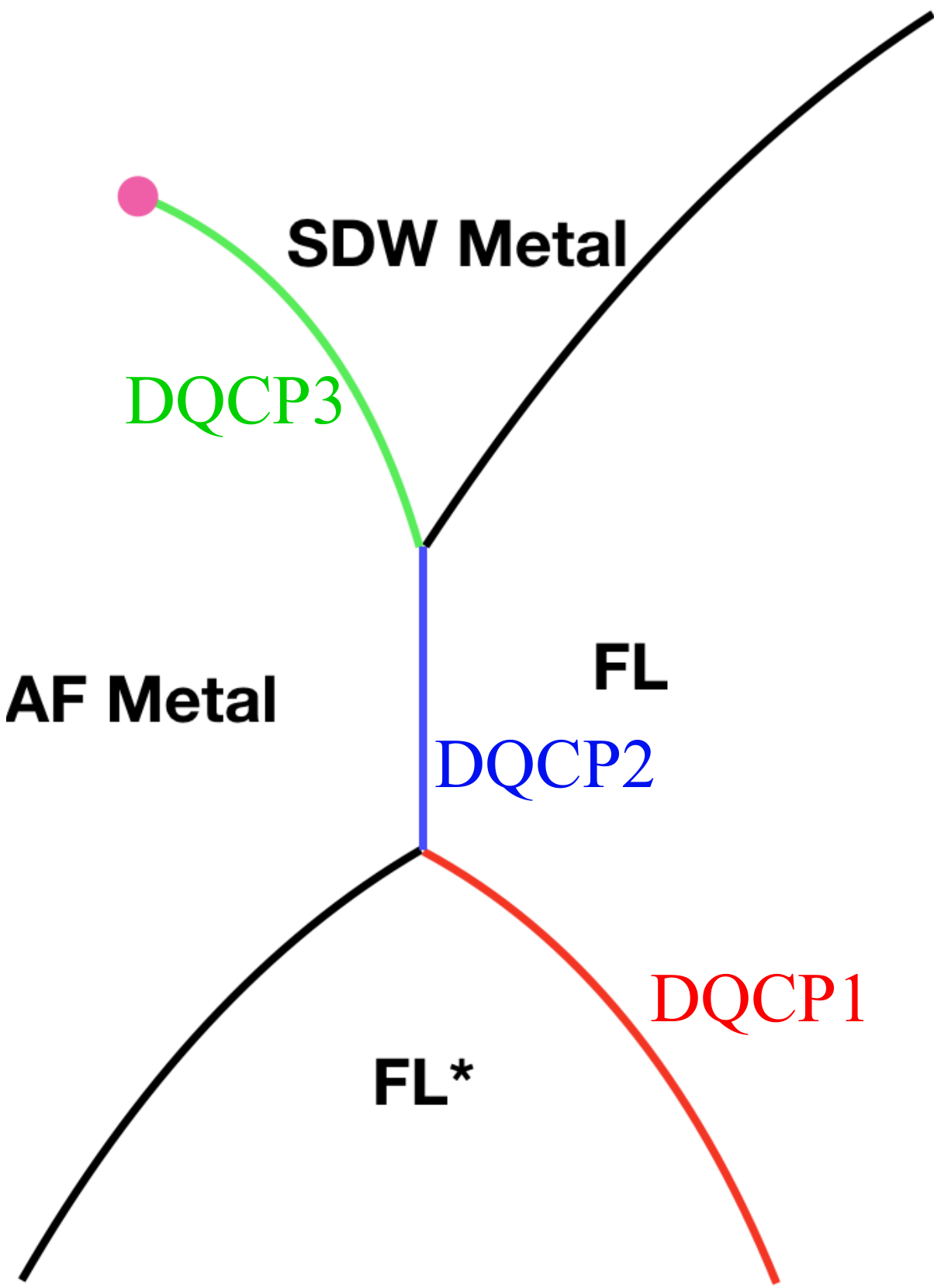}
    \caption{A global phase diagram obtained from our ancilla qubit approach. The red, blue, green lines are three different deconfined critical theories described in the present paper: DQCP1, DQCP2 and DQCP3.  DQCP1 is the FL*-FL transition. DQCP2 is the transition between AF Metal and the symmetric FL. DQCP3 is a transition within AF Metal. The DQCP3 may have an end point (denoted as a  solid circle) after which the transition becomes a crossover. This is a multicritical point corresponding to a Lifshitz transition point of the ghost Fermi surface in our theory (see Appendix~\ref{app:hm}). The FL-SDW Metal transition in a conventional transition in the Hertz-Millis class \cite{hertz,millis}, and the FL*-AF Metal transition is described by a QED$_3$ model discussed briefly in Section~\ref{sec:AFmetalmft}. At DQCP2, Kondo breakdown and onset of AF order coincide, while they are detached in opposite directions in DQCP1 and DQCP3.  Such a separation of Kondo breakdown and onset of magnetic order has been observed experimentally \cite{friedemann2009detaching}.  Therefore, all of these three DQCPs are potentially relevant to real experiments in heavy fermion systems. }    \label{fig:global_phase_diagram}
\end{figure}
We will present our results in the context of the global phase diagram presented in Fig.~\ref{fig:global_phase_diagram}. In addition to the FL and FL* phases noted above, it shows two types of metals with antiferromagnetic (N\'eel) order: the AF Metal and the SDW Metal. These are both `conventional' phases without excitations carrying charges of emergent gauge fields, and there is no fundamental distinction between them. Nevertheless the underlying physical interpretations of these phases are quite distinct. The SDW Metal arises from the appearance of a weak spin density wave in a FL state, and the consequent reconstruction of the Fermi surface.
On the other hand, the AF Metal can be considered as arising from a `Kondo breakdown' transition \cite{SiNature,ColemanSi01,SVS04}, where local moments appear due to absence of Kondo screening in the environment. These two pictures can lead to different Fermi surface shapes and topologies, but (apart from Fermi surface reconnections) the SDW Metal and AF Metal can be continuously connected, as indicated in Fig.~\ref{fig:global_phase_diagram}. 

Our analysis will lead to a sharper delineation of the distinction between the SDW Metal and AF Metal phases. As indicated in Fig.~\ref{fig:global_phase_diagram} the universality class of the phase transition from the FL to the AF Metal is distinct from the universality class of the transition from the FL to the SDW Metal: we will show that the first is described by a deconfined $(U(1)_S \times U(1)_1)/{Z_2}$ gauge theory, while the latter is in the conventional Hertz-Millis class \cite{hertz,millis}.   Another example of deconfined critical point for Landau symmetry breaking transition has been proposed in Ref.~\onlinecite{BiSenthil2}, but it does not involve a Fermi surface. Another surprising result in Fig.~\ref{fig:global_phase_diagram} is the presence of a sharp phase transition between the SDW Metal and the AF Metal, over a certain range of parameters. This is an example of an `unnecessary' quantum phase transition \cite{BiSenthil}, and will be described in our case by the deconfined $(U(1)_S \times U(1)_1)/Z_2$ gauge theory.

A crucial ingredient of the deconfined critical theories in Fig.~\ref{fig:global_phase_diagram} is a Fermi surface of ghost fermions coupled to 2 distinct gauge fields. The `S' gauge field ($SU(2)_S$ or $U(1)_S$) induces an attractive interaction between the ghost fermions, while the `1' gauge field ($U(1)_1$) induces a repulsive interaction. The ultimate fate of this critical ghost Fermi surface at very low temperatures depends on the delicate balance between these gauge forces, which we will describe in some detail following the analysis in Ref.~\onlinecite{MMSS14}. If the attractive forces dominate, then we obtain a superconducting state or a first-order transition at the lowest temperatures. If the repulsive forces dominate, then we obtain an intermediate non-Fermi liquid phase with a critical Fermi surface. We will see that attractive forces are likely to be stronger for DQCP1, while there is a near (DQCP2) or exact (DQCP3) balance between the leading gauge forces in the other cases, leading to exponentially small temperatures for possible instabilities. We note that weak instabilities are known to be an issue also for the canonical deconfined critical theory of antiferromagnets \cite{senthil1,senthil2}, but numerous numerical studies nevertheless exhibit the deconfined critical behavior over all accessible scales \cite{Zaletel20,Sreejith:2018ypg}.

We will begin in Section~\ref{sec:ancilla} with a general description of the ancilla approach, 
and the gauge theories it leads to.
The mean field structure of the phases of Fig.~\ref{fig:global_phase_diagram} will be described in Section~\ref{sec:MFT}. Here we will use a single band model for the physical layer. Section~\ref{sec:Kondo} will extend the mean field theory to a Kondo lattice in the physical layer, leading to very similar results.
We will begin description of the 3 DQCPs in Fig.~\ref{fig:global_phase_diagram} in Section~\ref{sec:crit}, by a
description of their overall and gauge and symmetry structures. The critical theories have three sectors of matter fields: a bosonic critical Higgs sector, a ghost Fermi surface coupled to different gauge fields, and a Fermi surface of gauge-neutral electrons in the physical layer.
We will first consider the critical ghost Fermi surface sector in Section~\ref{sec:critfermi}.
This will be followed by a consideration of the Higgs sector in Section~\ref{sec:higgs_boson},
and the physical Fermi surface of electrons in Section~\ref{sec:physical}. We will find that the electron Fermi surfaces exhibits marginal Fermi liquid behavior in certain regimes in three dimensions. Section~\ref{sec:conc} contains further physical discussion and an overview of our results (see Fig~\ref{fig:wavefunction} for a presentation of variational wavefunctions).

\section{Ancilla qubits and gauge symmetries}
\label{sec:ancilla}

We begin by reviewing basic aspects of the ancilla qubit approach \cite{Yahui1}, and the associated
$(SU(2)_S \times SU(2)_1 \times SU(2)_2)/Z_2$ gauge symmetry.

Let  $\vec{S}_{i;1}$, ${\bm S}_{i;2}$ be the spin operators acting on the qubits in the two hidden layers, where $i$ is a lattice site (see Fig.~\ref{fig:layers}). We can represent these spin operators with hidden fermions $F_{i;\sigma}$, $\widetilde F_{i;\sigma}$ via
\beq
{\bm S}_{i;1} = \frac{1}{2} F_{i;\sigma}^\dagger {\bm \sigma}_{\sigma\sigma'} F_{i ;\sigma'}^{}
\quad , \quad {\bm S}_{i;2} = \frac{1}{2} \widetilde F_{i;\sigma}^\dagger {\bm \sigma}_{\sigma\sigma'} \widetilde F_{i ;\sigma'}^{}
\label{eq:fermion_parton_spin}
\eeq
where ${\bm \sigma}$ are the Pauli matrices. Let us also define the Nambu pseudospin operators
\bea
{\bm T}_{i;1} &=& \frac{1}{2} \left(
F_{i;\downarrow}^\dagger F_{i;\uparrow}^\dagger + F_{i;\uparrow} F_{i;\downarrow} ,
i\left(F_{i;\downarrow}^\dagger F_{i;\uparrow}^\dagger - F_{i;\uparrow} F_{i;\downarrow} \right), 
F_{i;\uparrow}^\dagger F_{i;\uparrow} + F_{i;\downarrow}^\dagger F_{i;\downarrow} -1 \right) \nonumber \\
{\bm T}_{i;2} &=& \frac{1}{2} \left(
\widetilde F_{i;\downarrow}^\dagger \widetilde F_{i;\uparrow}^\dagger + \widetilde F_{i;\uparrow} \widetilde F_{i;\downarrow} ,
i\left(\widetilde F_{i;\downarrow}^\dagger \widetilde F_{i;\uparrow}^\dagger - \widetilde F_{i;\uparrow} \widetilde F_{i;\downarrow} \right), 
\widetilde F_{i;\uparrow}^\dagger \widetilde F_{i;\uparrow} + \widetilde F_{i;\downarrow}^\dagger \widetilde F_{i;\downarrow} -1 \right) \,.
\eea
For a more transparent presentation of the symmetries, it is useful to write the fermions as $2 \times 2$ matrices
\beq
\bm{F}_{i} = \left(
\begin{array}{cc}
F_{i ;\uparrow} & - F_{i ;\downarrow}^\dagger \\
F_{i ;\downarrow} & F_{i ;\uparrow}^\dagger
\end{array}
\right) \,.\label{Xc}
\eeq
This matrix obeys the relation
\beq
{\bm F}_{i}^\dagger = \sigma^y {\bm F}_{i}^T \sigma^y.
\eeq
We use a similar representation for $\widetilde{\bm F}$.
Now we can write the spin and Nambu pseudospin operators as
\bea
{\bm S}_{i;1} &=& \frac{1}{4} \mbox{Tr} ( {\bm F}_{i}^\dagger {\bm \sigma} {\bm F}_{i} ) \nonumber \\
{\bm T}_{i;1} &=& \frac{1}{4} \mbox{Tr} ( {\bm F}_{i}^\dagger {\bm F}_{i}  {\bm \sigma} )\,,
\eea
and similarly for ${\bm S}_{i;2}$ and ${\bm T}_{i;2}$ with $\widetilde{\bm F}$.

We now describe the gauge symmetries obtained by transforming to rotating reference frames in both spin and Nambu pseudospin spaces \cite{LeeWenRMP,SS09,XS10}. 
Note that in the ancilla approach \cite{Yahui1}, the gauge 
symmetries act only on the hidden layers, and we leave the degrees of freedom in the physical layer intact and gauge-invariant. The ${\bm F}$ and $\widetilde{\bm F}$ fermions already carry gauge charges associated with 
the $SU(2)$ gauge symmetries deployed in Ref.~\onlinecite{LeeWenRMP}, which we denote here as $SU(2)_1$ and $SU(2)_2$ for the two layers. These gauge symmetries can be associated with a transformation into a rotating reference frame in Nambu pseudospin space \cite{MSSS18}. We wish to work in the gauge invariant sector. So, as in Ref.~\onlinecite{LeeWenRMP} for the physical layer, we impose constraints on the hidden layers,
\beq
{\bm T}_{i;1} = 0 \quad , \quad {\bm T}_{i;2} = 0\,, \label{Tconstraints}
\eeq
which restrict each site of the hidden layers to single occupancy of the $F,\widetilde F$ fermions.

For the remaining gauge symmetries, we transform to rotating reference frame in spin space by introducing the `rotated' gauge-charged fermions $\bm{\Psi}_{i}$, $\widetilde{\bm \Psi}_i$ in the hidden layers via
\beq
{\bm F}_{i} = L_{i} \bm{\Psi}_{i}  \quad , \quad  \widetilde{\bm F}_{i} = \widetilde L_{i} \widetilde{\bm \Psi}_{i}   \label{FrotatePsi}
\eeq
where $L$, and $\widetilde L$ are $2 \times 2~$ $SU(2)$ matrices, and the $\bm{\Psi}$ fermions have a decomposition similar to 
(\ref{Xc})
\beq
{\bm \Psi}_{i} = \left(
\begin{array}{cc}
\Psi_{i ;+} & - \Psi_{i ;-}^\dagger \\
\Psi_{i ;-} & \Psi_{i ;+}^\dagger
\end{array}
\right) \,.\label{Xc1}
\eeq
We use indices $a=+,-$ for $\Psi_{i;a}$ rather than $\uparrow,\downarrow$ in (\ref{Xc1}) because the indices are not physical spin in the rotated reference frame. Again an analogous representation for $\widetilde \Psi_{ia}$ is used.
The transformation (\ref{FrotatePsi}) implies a rotation of the spin operators, but leaves the Nambu pseudospin invariant (and correspondingly for ${S}_{i;2}^\alpha$ and ${T}_{i;2}^\alpha$)
\bea
S_{i;1}^\alpha &=& \mathcal{L}_{i}^{\alpha \beta} \, \frac{1}{4} \mbox{Tr} ( {\bm \Psi}_{i}^\dagger \tau^\beta {\bm \Psi}_{i} ) \label{spint} \\
T_{i;1}^\beta  &=&  \frac{1}{4} \mbox{Tr} ( {\bm \Psi}_{i}^\dagger {\bm \Psi}_{i}  {\tau}^\beta ) \label{pspint}
\eea
where $\alpha,\beta = x,y,z$ and $\tau^\beta$ are Pauli matrices; we are using $\tau^\beta$ rather than $\sigma^\beta$ here to signify that these matrices act on the rotated $a=+,-$ indices. As the pseudospin is invariant, the constraints (\ref{Tconstraints}) now imply single occupancy of the $\Psi,\widetilde \Psi$ fermions.
The $\mathcal{L}_{i}$ is $3 \times 3~$ $SO(3)$ rotation matrix corresponding to the $2 \times 2~$ $SU(2)$ rotations:
\bea
\mathcal{L}_{i}^{\alpha \beta} &=& \frac{1}{2} \mbox{Tr} \left( L_{i}^\dagger \sigma^\alpha L_{i} \tau^\beta \right)\,.
\eea

Before analyzing the consequences of the rotation (\ref{FrotatePsi}), 
it is useful to tabulate the action of the $SU(2)_1$ and $SU(2)_2$ symmetries generated by the Nambu pseudospin operators, which are unchanged by the rotations in (\ref{FrotatePsi}).  We drop the site index $i$, as it is common to all fields:
\bea
SU(2)_1: \quad {\bm \Psi} \rightarrow {\bm \Psi} U_1 \quad &,& \quad \widetilde{\bm \Psi} \rightarrow \widetilde{\bm \Psi} \nonumber \\
L \rightarrow L \quad &,& \quad  \widetilde L \rightarrow \widetilde L \nonumber \\
SU(2)_2:~~~ \quad {\bm \Psi} \rightarrow {\bm \Psi} \quad &,& \quad \widetilde {\bm \Psi} \rightarrow \widetilde {\bm \Psi} U_2 \nonumber \\
L \rightarrow L \quad &,& \quad  \widetilde L \rightarrow \widetilde L \,. \label{su212}
\eea
Here, the gauge transformations are the $SU(2)$ matrices $U_1$ and $U_2$ respectively. Note that (\ref{su212}) corresponds to right multiplication of ${\bm \Psi}$,$\widetilde{\bm \Psi}$, and so commute with the gauge transformations associated with the left multiplication in (\ref{FrotatePsi}).

In considering the gauge constraints associated with (\ref{FrotatePsi}), we do {\it not\/} wish to impose the analog of the constraints (\ref{Tconstraints}) in the spin sector, because we don't want vanishing spin on each site of both layers. Rather, we want to couple the layers into spin singlets for each $i$, corresponding to the $J_\perp \rightarrow \infty$ limit in Fig.~\ref{fig:layers}. This is achieved by the constraints
\beq
{\bm S}_{i;1} + {\bm S}_{i;2} = 0\,. \label{Sconstraints}
\eeq
We will impose the constraints (\ref{Sconstraints}) at a finite bare gauge coupling, because
at infinite coupling the hidden layers would just form rung singlets. We do want to allow for some virtual fluctuations into the triplet sector at each $i$; otherwise, the hidden layers would completely decouple 
from the physical layer at the outset. In contrast, (\ref{Tconstraints}) is imposed at an infinite bare gauge coupling \cite{LeeWenRMP}. In practice, the value of the bare gauge coupling makes little difference, because we will deal ultimately with the effective low energy gauge theory.

The mechanism for imposing (\ref{Sconstraints}) is straightforward. We transform to a common rotating frame in both layers by identifying
\bea
\widetilde L_{i} &=& L_{i}  \nonumber \\
\widetilde {\mathcal L}_{i} &=& \mathcal{L}_{i} \,.
\eea
Now we have only a single $SU(2)_S$ gauge symmetry, related to that in Refs.~\onlinecite{SS09,XS10,Sachdev:2018ddg}, and the analog of (\ref{su212}) is
\bea
SU(2)_S: \quad {\bm \Psi} \rightarrow U_S {\bm \Psi}  \quad &,& \quad \widetilde{\bm \Psi} \rightarrow U_S \widetilde{\bm \Psi} \nonumber \\
L \rightarrow L U_S^\dagger\,, &~& \label{su2S}
\eea
where $U_S$ is an $SU(2)$ matrix.

We will assume $\langle L_i \rangle=0$ in the whole phase diagram as we are interested in projecting the hidden layers into rung spin singlets. After that, the crucial transformations for the subsequent development are those of the fermions ${\bm \Psi}$, which we collect here: 
\bea
SU(2)_1&:& \quad {\bm \Psi} \rightarrow {\bm \Psi} U_1 \quad , \quad \widetilde {\bm \Psi} \rightarrow \widetilde{\bm \Psi} \nonumber \\
SU(2)_2&:& \quad {\bm \Psi} \rightarrow {\bm \Psi}  ~~~~\quad , \quad \widetilde{\bm \Psi} \rightarrow \widetilde{\bm \Psi} U_2 
\nonumber \\
SU(2)_S&:& \quad {\bm \Psi} \rightarrow U_S {\bm \Psi}  \quad , \quad \widetilde{\bm \Psi} \rightarrow U_S \widetilde{\bm \Psi} \,. \label{gaugeall}
\eea
The reader need only keep track of (\ref{gaugeall}) for the following sections: the structure of all our effective actions is mainly dictated by the requirements of the gauge symmetries acting on the fermions in (\ref{gaugeall}), and on the Higgs fields that will appear in the different cases.
The $Z_2$ divisor in the overall $(SU(2)_1 \times SU(2)_2 \times SU(2)_S)/Z_2$ gauge symmetry arises from the fact that centers of the two $SU(2)$ tranformations in (\ref{FrotatePsi}) are the same.

\section{Mean field theories}
\label{sec:MFT}

We will begin by presenting some simple mean field theories of the $(SU(2)_S \times SU(2)_1 \times SU(2)_2)/Z_2$ gauge in the context of a one band model for the Hamiltonian of the physical layer. We will consider the extension to Kondo lattice models in Section~\ref{sec:Kondo}, and find that the phenomenology remains essentially the same.

Our discussion will take place in the context of the schematic global phase diagram in Fig.~\ref{fig:global_phase_diagram}.
The general strategy will be to break the $(SU(2)_S \times SU(2)_1 \times SU(2)_2)/Z_2$ gauge symmetry by a judicious choice of Higgs fields, and then examine the fluctuations of the Higgs fields that become critical at the boundaries between the phases. 

\subsection{FL*}
\label{sec:FL*singleband}

We first recall \cite{Yahui1} the mean field theory for the FL* phase. As we will see below this FL* phase will act as a `parent' phase for the other phases in Fig.~\ref{fig:global_phase_diagram}.

We represent the electrons in the physical layer by $C_\sigma$, using a notation following the convention in Section~\ref{sec:ancilla}. As we noted earlier, we will not fractionalize $C_\sigma$ directly: so the $C_\sigma$ does not carry any emergent gauge charges, only the global charges of the electromagnetic $U(1)_{em}$ and the spin rotation $SU(2)$. 
The FL* phase described by the following schematic Hamiltonian
\begin{equation}
  H_\ast =H_C+H_\Psi+ H_{\widetilde \Psi} + \sum_i \left( C^\dagger_{i;\sigma} \Phi_{i;\sigma a} \Psi_{i;a} + \mbox{H.c.} \right) \,,
  \label{HFL*}
\end{equation}
where $i$ is a lattice site index, $\sigma = \uparrow, \downarrow$ is a physical spin index, and $a = +,-$ is a gauge $SU(2)_S$ index.
The Hamiltonian $H_C$ is a generic one-band Hamiltonian for the electrons $C_\sigma$ (a specific form appears in (\ref{HSDW}), with $N_0=0$ in the FL* state). The Hamiltonian for the first hidden layer, $H_\Psi$, 
is a spin liquid Hamiltonian which breaks $SU(2)_1$ down to $U(1)_1$; at its simplest this could be a free fermion Hamiltonian with a trivial projective symmetry group (PSG) so that the $\Psi$ fermions on their own form a Fermi surface which occupies half the Brillouin zone (because the $\Psi$ density is at half-filling, from (\ref{Tconstraints})). Similarly, the Hamiltonian for the second hidden layer, $H_{\widetilde\Psi}$, 
is a spin liquid Hamiltonian which breaks $SU(2)_2$ down to $U(1)_2$; however now we use the `staggered-flux' ansatz \cite{LeeWenRMP}, so that $\widetilde{\Psi}$ excitations are Dirac fermions. (Specific forms for $H_\Psi$ and $H_{\widetilde\Psi}$ appear later in (\ref{HAF}), with $M_{1,2}=0$ in the FL* state).

The crucial term in (\ref{HFL*}) is the Higgs field $\Phi_{\sigma a}$, which is a $2 \times 2$ complex matrix linking the physical electrons to the first hidden layer. We will find it convenient to represent this by as pair of complex doublet $\Phi_a$, with
\beq
\Phi_{+}  = \left( \begin{array}{c} \Phi_{\uparrow +} \\ \Phi_{\downarrow +} \end{array} \right)
 \quad , \quad \Phi_{-}  = \left( \begin{array}{c} \Phi_{\uparrow -} \\ \Phi_{\downarrow -} \end{array} \right)
\eeq
From this representation, and the form of (\ref{HFL*}), it is clear that the $\Phi_a$ transform as follows under the various symmetries
\begin{itemize}
\item $\Phi_+$ and $\Phi_-$ transform separately as fundamentals under the global SU(2) spin rotation.
\item $\Phi_a$ also carries the global electromagnetic $U(1)_{em}$ charge, with $\Phi_a \rightarrow e^{i \theta} \Phi_a$ and $C \rightarrow e^{i \theta} C$ under a global $U(1)_{em}$ rotation.
\item $\Phi_a$ transforms as a fundamental under the $SU(2)_S$ gauge transformation, $\Phi_a \rightarrow \Phi_b [U_S^\dagger]_{ba}$, while, as in (\ref{gaugeall}), $\Psi_a \rightarrow [U_S]_{ab} \Psi_b$ and $\widetilde\Psi_a \rightarrow [U_S]_{ab} \widetilde\Psi_b$.
\item $\Phi_a$ carries the $U(1)_1$ gauge charge, with  $\Phi_a \rightarrow e^{-i \vartheta_1} \Phi_a$, along with $\Psi_a \rightarrow e^{i \vartheta_1} \Psi_a$.
\item $\Phi_a$ is neutral under the $U(1)_2$ gauge charge, with  $\Phi_a \rightarrow \Phi_a$, while $\widetilde\Psi_a \rightarrow e^{i \vartheta_2} \widetilde\Psi_a$.
\end{itemize}

The FL* phase arises when $\Phi_a$ is condensed. Then, the above transformations make it clear that both $SU(2)_S$ and $U(1)_1$ are fully broken (or `higgsed'). The electromagnetic $U(1)_{em}$ is however preserved because there is no gauge-invariant operator carrying $U(1)_{em}$ charge which acquires an expectation value. The condensation of $\Phi_a$ effectively ties $U(1)_1$ to $U(1)_{em}$ (this is as in the usual `slave particle' theories).
Similarly, if we choose $\Phi_{\sigma a} \propto \delta_{\sigma a}$ spin rotation invariance is preserved,
and $a$ becomes like an effective global spin index.

The properties of such a FL* phase were discussed in some detail in Ref.~\onlinecite{Yahui1}. The $C$ and $\Psi$ fermions are hybridized to form small pockets with total Fermi surface volume $A_{FS}=p/2$, where the electron density in the physical layer is $1-p$.  It is quite natural to expect that only Fermi arcs are visible in ARPES measurement while the backside of the pocket is dominated by $\Psi$ and has very small spectral weight in terms of the physical electron $C$. Because the $SU(2)_S$ gauge field is locked to the external physical spin gauge field,  the other ghost fermion $\widetilde \Psi$ now becomes a neutral spinon and forms a $U(1)$ Dirac spin liquid.

We will consider phase transitions out of this FL* phase later in this paper. In particular, the FL*-FL 
transition (see Fig.~\ref{fig:global_phase_diagram}) is described by a theory of the vanishing of the Higgs condensate $\langle \Phi_a \rangle$ while spin rotation is preserved. This yields a critical theory, denoted DQCP1 in Fig.~\ref{fig:global_phase_diagram}, in which the key degrees of freedom are the bosons $\Phi_a$ and the $\Psi$ Fermi surface coupled to $SU(2)_S \times U(1)_1$ gauge fields; the structure of the effective action can be deduced from the symmetry and gauge transformations we have described above. We will find in Section~\ref{sec:critfermi} that such a $\Psi$ Fermi surface is unstable to pairing, and FL*-FL transition likely occurs via an intermediate phase. 

\subsection{AF Metal}
\label{sec:AFmetalmft}

Next, we consider the AF Metal phase in Fig.~\ref{fig:global_phase_diagram}. This is a phase in spin rotation invariance is broken by antiferromagnetic order, and all the gauge fields are confined. We wish to obtain this state via a `Kondo breakdown' transition in which the spins are liberated to form an antiferromagnet, rather than a spin density wave instability of an electronic Fermi surface; the latter is denoted SDW Metal in Fig.~\ref{fig:global_phase_diagram}, and will be treated in Section~\ref{sec:SDWmetal}. So we propose an effective Hamiltonian which perturbs the Hamiltonian $H_\ast$ of the FL* phase in (\ref{HFL*}). The key idea is that the driving force for the appearance of antiferromagnetism in an AF Metal is the breaking of $SU(2)_S$ rather than global spin rotation symmetry; as $SU(2)_S$ is tied to global spin in the FL* phase, spin rotation symmetry will be broken as a secondary consequence. So the Hamiltonians of the 2 hidden layers, {\it i.e.\/} the ghost fermions, are as follows:
\begin{eqnarray}
H_{\Psi} &=& -t_{\Psi}\sum_{\langle ij \rangle} \Psi^\dagger_i \Psi_j-t'_{\Psi}\sum_{\langle \langle ij \rangle \rangle}\Psi^\dagger_i \Psi_j-M_1 \sum_i (-1)^i \Psi^\dagger_i \tau^z \Psi_i \nonumber \\ 
    H_{\widetilde \Psi} &=& -t_{\widetilde \Psi} \sum_{\langle ij \rangle} \widetilde \Psi^\dagger_i \widetilde \Psi_j+M_2 \sum_i (-1)^i \widetilde \Psi^\dagger_i \tau^z \widetilde \Psi_i \label{HAF}
\end{eqnarray}
As in Section~\ref{sec:ancilla}, we use $\tau^\alpha$ ($\alpha = x,y,z$) to represent Pauli matrices acting the $SU(2)_S$ gauge symmetry space; we use $\sigma^\alpha$ to represent Pauli matrices acting of the global spin rotation space.
The new ingredients in (\ref{HAF}), not present in the FL* phase, are the Higgs fields $M_{1,2}$ which break $SU(2)_S$ gauge symmetry down to $U(1)_S$. 
The presence of $M_2$ will gap out the $\widetilde{\Psi}$ Dirac fermions, and so $U(1)_2$ will confine. Recall that the $\langle \Phi_a \rangle$ condensate has already completely broken the $SU(2)_S$ and $U(1)_1$ gauge symmetries. So there are no remaining free gauge charges, and we obtain a conventional phase with global $SU(2)$ spin rotation symmetry broken down to $U(1)$. 

In considering the phase transition from the AF Metal to the FL phase in Fig.~\ref{fig:global_phase_diagram}
(labeled DQCP2), we consider the vanishing of the Higgs field $\Phi_a$, while $M_{1,2}$ remain non-zero. Then we obtain a critical theory which is very similar to the FL*-FL theory discussed above, except that the gauge symmetry is only $U(1)_S \times U(1)_1$ {\it i.e.\/} the main ingredients are the bosons $\Phi_a$ and the $\Psi_a$ Fermi surface coupled to $U(1)_S \times U(1)_1$ gauge fields. Note, however, that the global spin rotation invariance will be restored at the critical point, because spin rotation was only broken via the $\Phi_a$ condensate. 

We will show in Section~\ref{sec:critfermi} that this AF Metal-FL theory has an important difference from the FL*-FL theory: it is now possible to have a stable critical $\Psi_a$ Fermi surface which does not pair. Once we move to the other side of the critical point, and the Higgs fields are gapped, then the pairing instability does set in, and we expect a crossover to a confining phase which is likely to be a conventional FL state.

Let us also consider the FL*-AF Metal transition shown in Fig.~\ref{fig:global_phase_diagram}. Starting from the FL* state, this transition is realized by turning on a $M_2$ condensate, while the $\Phi_a$ condensate remains non-zero. Then, within the second hidden layer, the critical properties are described by a model considered earlier \cite{PouyanSenthil,Maciejko18,Janssen:2020cut,Dupuis:2019xdo}: an O(3) QED$_3$ Gross-Neveu-Yukawa model. The spectator FL* Fermi surfaces could have a significant influence on this conformal field theory, but we will not explore this here. We also note other approaches to the FL*-AF Metal transition \cite{TGTS10,Kaul08}, with different spin liquid structures in the FL* phase. 

\subsection{SDW Metal}
\label{sec:SDWmetal}

Now we consider the SDW Metal of Fig.~\ref{fig:global_phase_diagram}: this is conventional SDW Metal, obtained as an instability of the large Fermi surface of the electrons $C_\sigma$. So now we directly break the spin rotation symmetry by a $C_\sigma$ bilinear, in contrast to the indirect breaking via the ghost fermion bilinears in (\ref{HAF}) for the AF Metal. So the Hamiltonian $H_C$ in $H_\ast$ becomes
\beq
H_C = - \sum_{i,j} t_{ij} C^\dagger_i C^{}_j + N_0 \sum_i (-1)^i C^\dagger_i \sigma^z C^{}_j
\label{HSDW}
\eeq
where $N_0$ is proportional to the physical antiferromagnetic order. We note that as long as there is a $\Phi_a$ Higgs condensate, there is no fundamental distinction between the $N_0$ symmetry breaking in (\ref{HSDW}), and the 
$M_{1,2}$ Higgs condensates of the AF Metal in (\ref{HAF}). Both eventually lead to the same phase: a `trivial' phase with no emergent gauge charges, and long-range antiferromagnetic order. The distinction is only a question of degree. In the SDW Metal, $N_0$ is the primary driving force, and drives the appearance of the $M_{1,2}$ condensates as a secondary consequence, while the opposite is true in the AF Metal.
Nevertheless, there can be significant observable differences in the shapes and topologies of the Fermi surfaces, with those of the SDW Metal arising from a reconstruction of the large Fermi surface of the $C$ fermions.

Furthermore, we will show that there can be a novel `unnecessary phase transition' \cite{BiSenthil,BiSenthil2} between the AF Metal and the SDW Metal, denoted DQCP3 in Fig.~\ref{fig:global_phase_diagram}. The critical theory for this transition is very similar to that of the AF Metal-FL transition discussed above. The only distinction is that the global spin rotation symmetry is reduced from SU(2) to U(1): so we again obtain a theory of bosons $\Phi_a$ and the $\Psi$ Fermi surface coupled to $U(1)_S \times U(1)_1$ gauge fields.

\section{Kondo lattice}
\label{sec:Kondo}
Before turning to a consideration of the deconfined phase transitions in Fig.~\ref{fig:global_phase_diagram}, this section will extend the mean field considerations to the case where the physical layer is a Kondo lattice model.
The microscopic model was already illustrated in Fig.~\ref{fig:layers}; we consider
\bea
  H &=& H_K+H_{1}+H_2 +H_{S1}+H_{12} \nonumber \\
  H_K &=& -t\sum_{\langle ij \rangle} C^\dagger_i C_j +J_K \sum_i (C^\dagger_i \vec \sigma C_i) \cdot \vec S_i +J\sum_{\langle ij \rangle}\vec S_i \cdot \vec S_j \nonumber \\
H_1 &=& J_1 \sum_{\langle ij \rangle}\vec S_{j;1} \cdot \vec S_{j;1} \quad,\quad H_2=J_2\sum_{\langle ij \rangle}\vec S_{i;2} \cdot \vec S_{j;2} \nonumber \\
H_{12}& =& J_\perp \sum_i \vec S_{i;1} \cdot \vec S_{i;2} \quad, \quad H_{S1}=J_d \sum_i \vec S_i \cdot \vec S_{i;1}
\eea
Above $C_{i \sigma}$ is the itinerant electron, $\vec S_i$ is the local moment;  $\vec S_{i;1}$ and $\vec S_{i;2}$ are the spins in the hidden layers. We will take $J_\perp \rightarrow \infty$ limit to recover the standard Kondo-Heisenberg model.   

We perform the conventional heavy-fermion mean field theory on the physical layer, and the parton description described in Section~\ref{sec:ancilla} on the hidden layers. 
The physical layer is represented by a 2-band model involving hybridization between the electrons $C_{i \alpha}$ and the $f_{i \sigma}$ fermions used to represent the Kondo lattice spins ${\bf S}_{i}$, and 
the hidden layers are represented by the mean field theory for $\Psi$,$\widetilde{\Psi}$ described in Section~\ref{sec:MFT}. 
So, our mean field theory is:
\bea
  H_M & =& H_{c,f}+H_{\Psi}+H_{\rm Higgs}+H_{\widetilde \Psi} \nonumber \\
  H_{c,f}&=& -t_c\sum_{\langle ij \rangle}C^\dagger_i C_j -t_f \sum_{\langle ij \rangle}f^\dagger_i f_j
  -t'_f \sum_{\langle \langle ij \rangle \rangle} f^\dagger_i f_j -t''_f \sum_{\langle \langle \langle ij \rangle \rangle \rangle} f^\dagger_i f_j +K \sum_i \left(C^\dagger_i f_i + \mbox{H.c.} \right) \,,
\eea
and $H_\Psi$ and $H_{\widetilde \Psi}$ are as in (\ref{HAF}).
Finally we need to add the Higgs term to describe the Kondo breaking down
\begin{equation}
  H_{\rm Higgs}= \sum_i \left( f^\dagger_{i;\sigma} \Phi_{i;\sigma a} \Psi_{i;a} + \mbox{H.c.} \right)\,,
\end{equation}
similar to that (\ref{HFL*}). If both $M_{1,2}$ and $\Phi_a$ are non-zero, then the physical spin rotation symmetry is broken, and we obtain a AF Metal state with N\'eel order.

We always assume the Kondo coupling $K\neq 0$, and therefore $f$ can also be viewed as physical electron.  
The Kondo breakdown happens when we add $\Phi \neq 0$. (This is to be compared with a theory of the FL*-FL transition in a Kondo lattice model \cite{SSV03,SVS04}, where Kondo breakdown occured via the vanishing of the Higgs field representing the Kondo coupling $K$.) With large $\Phi$, $f$ will be tightly bound with $\Psi$ and both disappear in the low energy limit. This is exactly what we expect from the picture that $f$ gets Mott localized.  In the low energy limit there is still a degree of freedom from $\widetilde \Psi$, which can now be viewed as a spinon for the localized moment after Mott localization.  Because of the ansatz we assumed for $H_{\widetilde \Psi}$, these local moments form a Neel order.   In summary, the final state is a small pocket formed by $C$ moving on top of Neel order formed by localized moment.   This picture is different from a SDW Metal, although there is no sharp distinction in terms of symmetry and topology.

\begin{figure}
\centering
\includegraphics[width=0.95\textwidth]{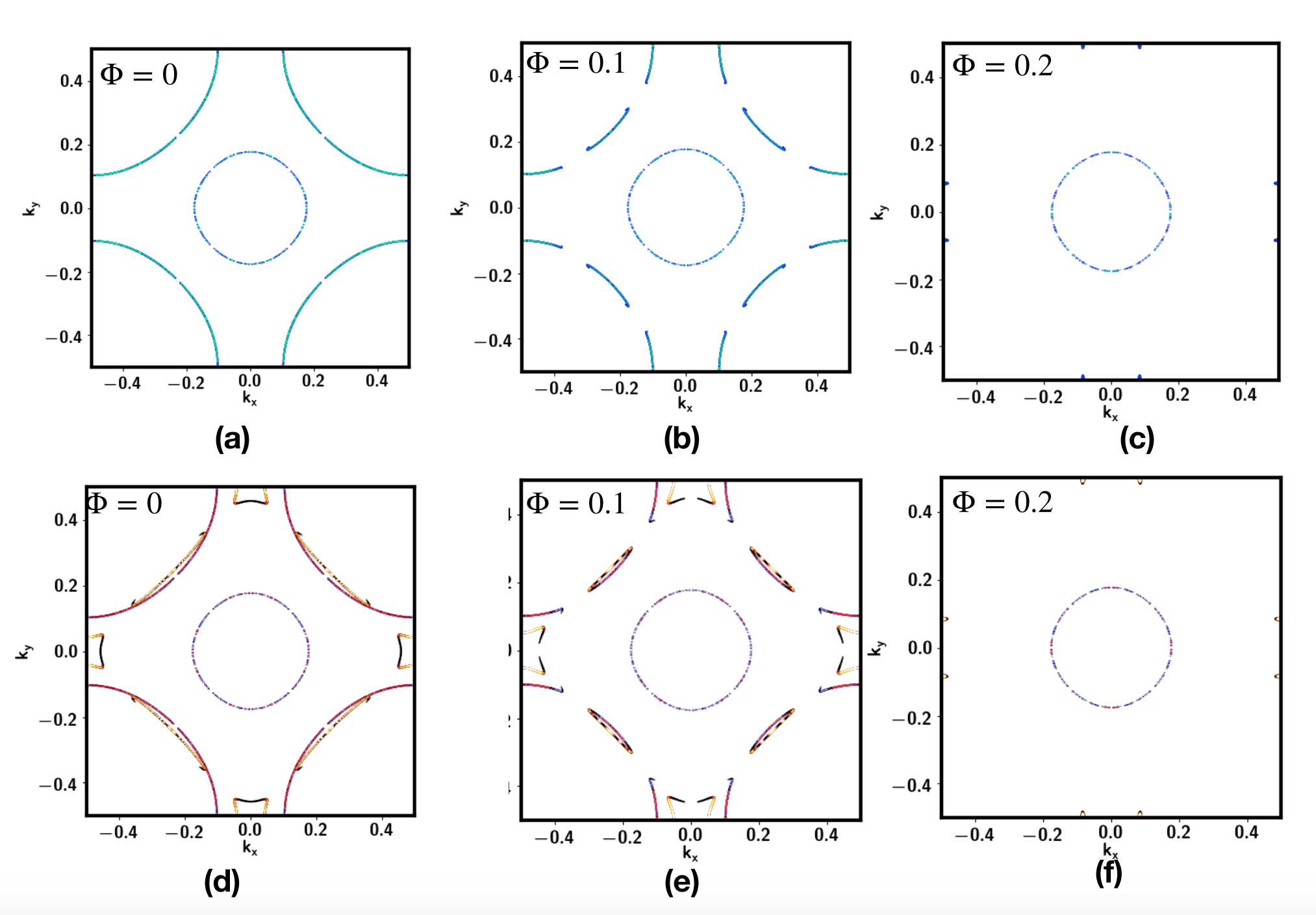}
\caption{Evolution of the spectral function $A(\omega=0,\mathbf k)=\text{Im} G(\omega,\mathbf k)$ when increasing $\Phi$ for the AF metal phase. Here $G(\omega,\mathbf k)$ is the Green function. In the first row we plot $A_c(\omega=0,\mathbf k)+A_f(\omega=0,\mathbf k)$. In the second row we further include the contribution from $\Psi$ and plot $A_c(\omega=0,\mathbf k)+A_f(\omega=0,\mathbf k)+A_\Psi(\omega=0,\mathbf k)$. $k_x$ and $k_y$ are in unit of ${2\pi}/{a}$, where $a$ is the lattice constant.}
\label{fig:apres_AF_metal}
\end{figure}

In Fig.~\ref{fig:apres_AF_metal} we show plots of the single electron spectrum after the transition of condensing $\langle \Phi_{\sigma a} \rangle=\Phi \delta_{\sigma a}$.  We use $t_c=10$, $t_f=1$, $t'_f=-0.1$, $t''_f=0.1$,  $t_\psi=-1$ and $t'_\psi=0.1$. In our framework the Kondo coupling $K$ is finite at the Kondo breakdown transition. But it is natural to expect that $K$ is small at the quantum critical point, thus we use $K=1$.  At this value, the heavy Fermi liquid has two separate pockets: one small Fermi surface mainly from $c$ and one large one mainly from $f$, as shown in Fig.~\ref{fig:apres_AF_metal}(a).  When we add a small value of $\Phi$, $f$ and $\Psi$ get hybridized and gapped after a small inter-mediate region.  Here only $\Phi=0.2$ is needed to fully gap out the Fermi surface formed by $f$.  In our convention the hopping of the spinon is $1$, which implies the energy scale of the spin-spin interaction is at order of $1-10$.  Thus $\Phi=0.2$ is only a few percent of the spin-spin interaction scale.  The transition we are describing is a Kondo breaking down transition and $\Phi$ is associated with the charge gap and should be determined by Hubbard $U$.  Therefore the region $\Phi<0.2$ should be very small and in the real experiment one may see an almost abrupt destruction of the Fermi surface associated with $f$.

\section{Structure of Critical Theories}
\label{sec:crit}

Our discussion in Section~\ref{sec:MFT} of the phases of Fig.~\ref{fig:global_phase_diagram}, also stated the ingredients of the effective theory of the critical points between them. All critical theories have the same important matter degrees of freedom: a 4-component complex scalar $\Phi_{a \sigma}$, and a Fermi surface of the 2-component complex fermions $\Psi_{a}$ at half-filling. Both matter fields carry fundamental gauge charges, and are subject to different symmetry constraints. In summary, these are:
\begin{itemize}
    \item DQCP1 between FL and FL* is described by a $(SU(2)_S \times U(1)_1)/Z_2$ gauge theory with a global SU(2) spin rotation symmetry.
    \item DQCP2 between FL and AF Metal is described by a $(U(1)_S \times U(1)_1)/Z_2$ gauge theory with a global SU(2) spin rotation symmetry.
    \item DQCP3 between AF Metal and SDW Metal is described by a $(U(1)_S \times U(1)_1)/Z_2$ gauge theory with a global U(1) spin rotation symmetry.
\end{itemize}
In addition to these gauge and global symmetries, lattice and time reversal symmetries also play an important role. We discuss these, and the resulting effective actions for the fermions and bosons in the following subsections.

Generically, the Lagrangian for the critical theory has the following structure:
\begin{equation}
    L=L_C+L_{\Psi}+L_{\Phi}+g\sum_{a=\pm}(C^\dagger \Phi_a) \Psi_{a} +L_{\Phi \Psi}+L_{\Phi C} \label{global}
\end{equation}
where $L_C$ is the Lagrangian for the Fermi surface of $C$ electrons in the physical layer, $L_{\Phi}$ is the Lagrangian for the Higgs field, and $L_{\Psi}$ is the Lagrangian for the ghost fermion sector. Both $\Phi$ and $\Psi$ couple to the internal gauge fields and we have suppressed the action for the gauge fields.  The main communication between the physical sector $C$ and the hidden sector $\Phi,\Psi$ is through the Yukawa coupling $g$ in the above, but there are also terms in the form $(C^\dagger \sigma^\beta C)(\Phi^\dagger \tau^\alpha \sigma^\beta \Phi)$ included in $L_{\Phi C}$.

\subsection{Lattice symmetry}
\label{sec:lattice}

The projective lattice symmetry plays an essential role in the critical theory for DQCP2 and DQCP3. Therefore we list the lattice symmetries first for the three DQCPs. We will only focus on the time reversal symmetry $T$ and translation symmetry $T_x,T_y$.

\subsubsection{DQCP1}

For DQCP1, the lattice symmetry transforms trivially.
\bea
T&:& \quad{ C} \rightarrow i\sigma^y  C, \quad { \Psi} \rightarrow { \Psi}, \quad { \Phi} \rightarrow  i\sigma^y { \Phi} \nonumber \\
T_x&:& \quad{ C} \rightarrow  C, \quad { \Psi} \rightarrow { \Psi}, \quad { \Phi} \rightarrow  { \Phi} \nonumber \\
T_y&:& \quad{ C} \rightarrow  C, \quad { \Psi} \rightarrow { \Psi}, \quad { \Phi} \rightarrow  { \Phi} \label{lattice_symmetry_DQCP1}
\eea
Here we suppress the transformation of the coordinate $(x,y)$ of the field. $T$ always transforms $(x,y)$ to $(x,y)$.  $T_x$ transforms $(x,y)$ to $(x+1,y)$ and $T_y$ transforms $(x,y)$ to $(x,y+1)$.

\subsubsection{DQCP2}

For DQCP2, the translation symmetry needs to act projectively to keep the ansatz (\ref{HAF}) invariant. Because of the $M_1$ term, we need to add a gauge transformation $i\tau^y \in SU(2)_S$ after $T_x$ and $T_y$.
\bea
T&:& \quad{ C} \rightarrow i\sigma^y  C, \quad { \Psi} \rightarrow { \Psi}, \quad { \Phi} \rightarrow  i\sigma^y { \Phi} \nonumber \\
T_x&:& \quad{ C} \rightarrow  C, \quad { \Psi} \rightarrow i\tau^y { \Psi}, \quad { \Phi} \rightarrow  i\tau^y{ \Phi} \nonumber \\
T_y&:& \quad{ C} \rightarrow  C, \quad { \Psi} \rightarrow i\tau^y { \Psi}, \quad { \Phi} \rightarrow  i\tau^y{ \Phi} \label{lattice_symmetry_DQCP2}
\eea

\subsubsection{DQCP3}
For DQCP3, there is always a Neel order parameter as shown in (\ref{HSDW}). Both the time reversal symmetry and the translation symmetry are already broken. We are left with the combined symmetry $T_xT$ and $T_yT$.
\bea
T_x T&:& \quad{ C} \rightarrow i\sigma^y  C, \quad { \Psi} \rightarrow i\tau^y { \Psi}, \quad { \Phi} \rightarrow  -\tau^y\sigma^y{ \Phi} \nonumber \\
T_y T&:& \quad{ C} \rightarrow i\sigma^y  C, \quad { \Psi} \rightarrow i\tau^y { \Psi}, \quad { \Phi} \rightarrow  -\tau^y\sigma^y{ \Phi} 
\label{lattice_symmetry_DQCP3}
\eea

\subsection{Action for ghost fermions $\Psi$}

Here we introduce the Lagrangian $L_{\Psi}$ in (\ref{global}).

\subsubsection{DQCP1}
For the FL-FL* transition, $\Psi$ couples to the $U(1)_1$ and $SU(2)_S$ gauge fields. We can ignore the $\widetilde \Psi$ part, and the $U(1)_2$ gauge field, which are not touched across the transition.
We label the $U(1)_1$ gauge field as $a_1$ and the $SU(2)_S$ gauge field as $a_S$. We have
\begin{equation}
    L_{\Psi}= \Psi^\dagger(\partial_\tau-i a_{1;0}-ia_{S;0}^\alpha \tau^\alpha) \Psi-\frac{\hbar^2}{2m_\Psi}  \Psi^\dagger (\partial_i-i a_{1;i}-ia_{S;i}^\alpha \tau^\alpha)^2 \Psi+...
\end{equation}
where $\alpha=x,y,z$ labels the three generators for the $SU(2)_S$ transformation. The ellipses denotes additional terms, including the four fermion interaction. Note that $\Psi$ does not carry either physical spin or physical charge. The lattice symmetry acts trivially in this case.

\subsubsection{DQCP2 and DQCP3}

For DQCP2 and DQCP3, $\Psi$ couples to a $(U(1)_1\times U(1)_S)/Z_2$ gauge field. We label $a_1$ and $a_S$ for $U(1)_1$ and $U(1)_S$ respectively.  It is convenient to relabel $a_+=a_1+a_S$ and $a_-=a_1-a_S$, then $\Psi_+$ couples to $a_+$ and $\Psi_-$ couples to $a_-$.  The low energy Lagrangian is of the form
\begin{equation}
    L_{\Psi}=\sum_{a=\pm}\Psi^\dagger_a(\partial_\tau-i a_{a;0})\Psi_a- \frac{\hbar^2}{2m_\Psi}\sum_{a=\pm}\Psi^\dagger_a(\partial_i-i a_{a;i})^2 \Psi_a+... \label{LPsi23}
\end{equation}
On the square lattice, the $M_1$ term in (\ref{HAF}) will reconstruct $\Psi_a$ to have electron and hole pockets with the same size (the sum over pockets is not indicated above). But the Fermi surfaces for $\Psi_+$ and $\Psi_-$ are always the same and inversion symmetric. Especially, there is perfect nesting for the pairing instability between $\Psi_+$ and $\Psi_-$.

\subsection{Action for Higgs bosons $\Phi$}
\label{sec:Higgsaction}

Next we turn to the sector for the Higgs bosons, and Lagrangian $L_\Phi$ in (\ref{global}), invariant under the symmetries in Section~\ref{sec:lattice}. In contrast to previous critical Higgs theories for the cuprates \cite{Chowdhury:2014efa,Sachdev:2018ddg,SSST19,SPSS20,SSS20} where the Higgs fields only carried gauge charges, here the Higgs fields also carry the physical electromagnetic and spin rotation quantum numbers. This has the important consequence that the Fermi surface induced Landau damping appears not in a term quadratic in the Higgs field (as in Refs.~\cite{Chowdhury:2014efa,Sachdev:2018ddg}), but in the quartic term in $L_\Phi'$ to be discussed in Section~\ref{sec:higgs_boson}.

\subsubsection{DQCP1}

The Higgs boson $\Phi$ is like an exciton formed by $C^\dagger$ and $\Psi$. As its density is not fixed, the condensation transition should have a term at linear order of $\partial_\tau$.
\begin{align}
  L_\Phi&= \Phi^\dagger(\partial_\tau-iA_0+i a_{1;0}+ia^{\alpha}_{S;0} \tau^\alpha)\Phi+\frac{1}{2m_\Phi}\Phi^\dagger(\partial_\mu-i A_\mu+i a_{1;\mu}+i a^{\alpha}_{S;\mu} \tau^\alpha)^2 \Phi+\mu_\Phi \Phi^\dagger  \Phi+\tilde \lambda( \Phi^\dagger  \Phi)^2.  \notag \\
  &+\tilde \lambda_1 ( \Phi^\dagger  \vec \sigma  \Phi)\cdot ( \Phi^\dagger  \vec{\sigma}  \Phi)+\tilde \lambda_2 ( \Phi^\dagger \vec\tau    \Phi)\cdot ( \Phi^\dagger \vec \tau   \Phi)+\tilde \lambda_3 \sum_{\alpha,\beta=x,y,z} ( \Phi^\dagger \tau^\alpha \sigma^\beta   \Phi)\cdot ( \Phi^\dagger  \tau^\alpha \sigma^\beta  \Phi) 
\end{align}
where $A_\mu$ is a background gauge field, introduced as a source for the global electromagnetic $U(1)_{em}$ charge. The linear time derivative term will also appear for DQCP2 and DQCP3, and consequently there will be some similarities between our critical theories and those studied previously for transitions studied previously for multi-band models and known variously as FL*-FL, Kondo breakdown, or orbitally selective Mott transitions \cite{SVS04,Pepin1,Pepin2,Pepin3}. However, there is also an important difference from these past studies: in the previous work the `hybridization-Higgs' boson condenses on the large Fermi surface side, while in our work it condenses on the small Fermi surface side (see Fig.~\ref{fig:wavefunction})

There are 4 quartic terms invariant under gauge symmetries and those in Section~\ref{sec:lattice}, but they are not all linearly independent of each other.
We can simplify the quartic terms by organizing them in the form $(\Phi^\dagger_{a\sigma_1} \Phi_{a\sigma_2})(\Phi^\dagger_{b\sigma_3} \Phi_{b\sigma_4})$,where $a,b=+,-$ and $\sigma_1,\sigma_2,\sigma_3,\sigma_4=\uparrow,\downarrow$. 
By using the identities 
\bea
(\Phi^\dagger \vec \sigma \Phi)\cdot (\Phi^\dagger \vec \sigma \Phi) &=& (\Phi^\dagger \vec \tau \Phi)(\Phi^\dagger \vec \tau \Phi) \nonumber \\
&= & 2(\Phi^\dagger_+ \vec \sigma \Phi_+) \cdot (\Phi^\dagger_- \vec \sigma \Phi_-)+\frac{1}{2}(\Phi^\dagger \Phi)^2+\frac{1}{2}(\Phi^\dagger \tau_z \Phi)^2 \nonumber \\ 
\sum_{\alpha,\beta=x,y,z} ( \Phi^\dagger \tau^\alpha \sigma^\beta   \Phi)\cdot ( \Phi^\dagger  \tau^\alpha \sigma^\beta  \Phi) &=& 2(\Phi^\dagger \Phi)^2-(\Phi^\dagger \tau_z \Phi)^2-4(\Phi^\dagger_+ \vec \sigma \Phi_+) \cdot (\Phi^\dagger_- \vec \sigma \Phi_-)\,, 
\eea
we find the Lagrangian can be written with only two independent quartic couplings $\lambda,\lambda_1$
\begin{align}
  L_\Phi&= \Phi^\dagger(\partial_\tau-iA_0+i a_{1;0}+ia^{\alpha}_{S;0} \tau^\alpha)\Phi+\frac{1}{2m_\Phi}\Phi^\dagger(\partial_\mu-i A_\mu+i a_{1;\mu}+i a^{\alpha}_{S;\mu} \tau^\alpha)^2 \Phi- \mu_\Phi \Phi^\dagger  \Phi \notag \\
  & ~~+\lambda( \Phi^\dagger  \Phi)^2+\lambda_1(\Phi^\dagger \tau_z \Phi)^2+4\lambda_1 (\Phi^\dagger_+ \vec \sigma \Phi_+) \cdot (\Phi^\dagger_- \vec \sigma \Phi_-)
   \label{eq:Higgs_boson_DQCP1}
\end{align}
It is easy to notice that $\vec n= \Phi^\dagger   \vec \sigma  \Phi$ is gauge invariant and represents spin fluctuation at momentum $\mathbf Q=(0,0)$.

\subsubsection{DQCP2}
For DQCP2, we have
\begin{align}
  L_\Phi&= \sum_{a=\pm }\Phi^\dagger_a(\partial_\tau-iA_0+ia_{a;0})\Phi_a+\frac{1}{2m_\Phi}\Phi^\dagger_a(\partial_\mu-i A_\mu+i a_{a;\mu})^2 \Phi_a
  -\mu_\Phi \sum_{a=\pm} |\Phi_a|^2\notag\\
  &+~~\lambda \sum_{a=\pm}|\Phi_a|^4 +
  \lambda_1 |\Phi_+|^2 |\Phi_-|^2+\lambda_2 (\Phi_+^\dagger\vec \sigma  \Phi_+) \cdot (\Phi_-^\dagger \vec \sigma \Phi_-) 
  \label{eq:Higgs_boson:DQCP2}
\end{align}
Because $(\Phi^\dagger_a \vec \sigma \Phi_a)\cdot (\Phi^\dagger_a \vec \sigma \Phi_a ) =(\Phi_a^\dagger \Phi_a)^2$, there are no other quartic terms with the $SU(2)$ global spin rotation symmetry.

The action of the symmetries in Section~\ref{sec:lattice} 
allows us to define a set of intertwined order parameters from the $\Phi_{a \sigma}$.
We can organize the order parameter as $O_{\alpha \beta}=\Phi^\dagger \tau^\alpha  \sigma^{\beta} \Phi$, 
where $\tau^\alpha$ are matrices in the gauged $SU(2)_S$ space with indices $a,b$, and $\sigma^\beta$ are matrices in the global SU(2) spin rotation space. To be gauge invariant, we need to restrict $\alpha=0,z$, but $\beta$ can be any one from $0,x,y,z$.
\begin{itemize}
    \item $O_{00}=\Phi^\dagger \Phi$ corresponds to density fluctuation.
    \item $O_{z0}=\Phi^\dagger \tau^z \Phi=\Phi_+^\dagger \Phi_+ -\Phi_-^\dagger \Phi_-$ is a CDW order parameter with $\mathbf Q=(\pi,\pi)$.
    \item $\vec n_i^{\text{AF}} \sim (-1)^i \vec{S}_i \sim \Phi^\dagger \tau^z \vec \sigma \Phi= (\Phi_+^\dagger \vec \sigma \Phi_+-\Phi_-^\dagger \vec \sigma \Phi_-)$  is the Neel order parameter.
    \item  $\vec n^{\text{FM}}_i=\Phi^\dagger \vec \sigma \Phi=(\Phi_+^\dagger \vec \sigma \Phi_++\Phi_-^\dagger \vec \sigma \Phi_-)$ is a ferromagnetic (FM) order parameter.
\end{itemize}
One can easily check that $T$ acts as $T: \vec n_i^{\text{AF}} \rightarrow - \vec n_i^{\text{AF}}$ and translation acts as $T_x: \vec n_i^{\text{AF}} \rightarrow -\vec n_{i+\hat{x}}^{\text{AF}}$. Meanwhile, $\vec n^{\text{FM}} \rightarrow -\vec n^{\text{FM}}$ under time reversal $T$ and $\vec n^{\text{FM}} \rightarrow \vec n^{\text{FM}}$ under translation.

Because several intertwined order breaking have algebraic decay at the QCP, the exact symmetry breaking pattern  at the $\mu_\Phi>0$ side  is selected by the quartic terms. For AF order, we need $\lambda_1<2\lambda$ and $\lambda_2>0$.

\subsubsection{DQCP3}

For DQCP3, the physical Neel order parameter will favor $\varphi_+=\Phi_{+;\uparrow}$ and $\varphi_-=\Phi_{-;\downarrow}$.  The $T_x T$ symmetry acts as $\varphi_+\rightarrow -\varphi_-, \varphi_-\rightarrow \varphi_+$.  The action reduces to:
\begin{align}
  L_\Phi&= \sum_{a=\pm }\varphi^\dagger_a(\partial_\tau-iA_0+ia_{a;0})\varphi_a+\frac{1}{2m_\Phi}\varphi^\dagger_a(\partial_\mu-i A_\mu+i a_{a;\mu})^2 \varphi_a
  -\mu_\Phi \sum_{a=\pm} |\varphi_a|^2\notag\\
  &+~~\lambda \sum_{a=\pm}|\varphi_a|^4+\lambda_1 |\varphi_1|^2 |\varphi_2|^2
  \label{eq:Higgs_boson_DQCP3}
\end{align}
The AF order parameter is now $n^{\text{AF}}_z=\varphi_+^\dagger \varphi_++\varphi_-^\dagger \varphi_-$, while the FM order parameter is $n^{\text{FM}}_z=\varphi_+^\dagger \varphi_+-\varphi_-^\dagger \varphi_-$. One can easily see that $\langle n^{\text{AF}}_z \rangle \neq 0$ for both $\mu_\Phi>0$ and $\mu_\Phi<0$, indicating a non-zero Neel order across the QCP.

\section{Instability of Critical ghost Fermi surfaces}
\label{sec:critfermi}

This section will consider the physics of the critical Fermi surfaces for the ghost Fermion $\Psi$ in the DQCP theories introduced in Section~\ref{sec:crit}. Our approach will be to consider an effective action for the gauge fields, and then compute the consequences of these gauge fields on the ghost fermions near the Fermi surface.

One important consequence of the form of the gauge field action  is an enhancement of the linear-in-$T$ specific heat of the Fermi surface. The ghost fermion surface has a $T^{{2}/{3}}$ 
specific heat in $2+1$ dimensions and $T\log(1/T)$ specific heat in $3+1$ dimensions  because the effective mass of $\Psi$ diverges due to the  gauge field \cite{SenthilMott,Mross:2010rd,MMSS14}. This appears to be 
compatible with experimental observations.

The DQCP2 and DQCP3 are actually critical lines with the parameter $M_1$ changing.  Upon increasing $M_1$, the ghost Fermi surface of $\Psi$ shrinks and finally gaps out through a 
Lifshitz transition. For DQCP2, when there are no ghost Fermi surfaces, the AF Metal-FL critical theory reduces to the Hertz-Millis \cite{hertz,millis} theory, as we show in Appendix~\ref{app:hm}. For DQCP3, after $\Psi$ is fully gapped out, there is no longer phase transition and we only expect a crossover. The red point in Fig.~\ref{fig:global_phase_diagram} is a multicritical point at the Lifshitz transition of the ghost Fermi surface. This demonstrates the crucial role played by the ghost Fermi surfaces in the deconfinement at the critical point.

The three theories presented in Section~\ref{sec:crit} are distinguished by the global spin rotation symmetry and the gauge fields. The global spin symmetry will not play any role in the present section. So the cases to consider are fermions coupled to $(SU(2)_S \times U(1)_1)/Z_2$ gauge fields, and to $(U(1)_S \times U(1)_1)/Z_2$ gauge fields. 
The ghost Fermi surface in DQCP1 with $(SU(2)_S \times U(1)_1)/Z_2$ gauge fields is argued to be strongly unstable to pairing at zero magnetic field in Appendix~\ref{append:RG_U2}.  We will only consider the $(U(1)_S \times U(1)_1)/Z_2$ case here (as it is relevant to the AF Metal-FL DQCP2 transition of central interest, and to the DQCP3 transition), and describe the $(SU(2)_S \times U(1)_1)/Z_2$ case in Appendix~\ref{append:RG_U2}. 

For the  $(U(1)_S \times U(1)_1)/Z_2$ gauge theory, we will show in the following subsections that there needs to be an intermediate phase (or first-order transition) between the AF metal and the FL phase. However, the scale of such ordering is exponentially suppressed, and the behavior above this small scale can be viewed as from a single deconfined critical point. Our analysis will based upon the renormalization group method of Ref.~\onlinecite{MMSS14}.

\subsection{Self-Energy of the photon}
\label{sec:diamagfermion}

We analyze the stability of the QCP at $\mu_\Phi=0$. In this case, after a renormalization of the gauge field Lagrangian from polarization corrections from the fermions $\Psi$ and bosons $\Phi$, we obtain \cite{MMSS14}:
\begin{equation}
  L_{a}=\frac{1}{2} \left(\frac{1}{e_{c;0}^{2}}{|\mathbf q|}^2 +\kappa_0 \frac{|\omega|}{|{\mathbf q}|} \right) |a_1(\omega,\mathbf q)|^2+ \frac{1}{2} \left(\frac{1}{e_{s;0}^{2}}|\mathbf q|^2+\kappa_0 \frac{|\omega|}{{\mathbf q}} \right) |a_S(\omega,\mathbf q)|^2 \label{Lacrit}
\end{equation}
Here the $|\omega|/|{\mathbf q}|$ terms are from Landau damping from the $\Psi$ Fermi surfaces. The $|{\mathbf q}|^2$ term for the gauge field should be mainly from the diagmagnetism of the ghost Fermion $\Psi$, and thus we expect $e_{c;0}^2\approx e_{s;0}^2 \approx {1}/{(\chi_f)}$, where $\chi_f$ is the diagmagnetism from one flavor of ghost fermion.  We have $\chi_f={1}/{ (12\pi m_{\Psi})} $ in $2+1$ dimensions.  

The difference between the coupling constants of the two gauge fields will play an important role below, and so we define a coupling which will play an important role below
\beq
r=\frac{e_{s;0}^2-e_{c;0}^2}{e^2_{c;0}} \,. \label{defr}
\eeq
To implement the gauge constraint exactly, we need the bare Maxwell term to be zero. Consequently, $e_{c,0}^2$ and $e_{s;0}^2$ arise from integration of $\Psi$ and $\Phi$.  We discuss the contributions from $\Psi$ and $\Phi$ respectively.

It is useful to rewrite the action for the gauge field in terms of $a_+=a_1+a_S$ and $a_-=a_1-a_S$:
\begin{equation}
    L_a= \Pi_{++}(\mathbf q) |a_+(\mathbf q)|^2+\Pi_{--}(\mathbf q)|a_-(\mathbf q)|^2+\Pi_{+-}(\mathbf q) \big(a_+(\mathbf q) a_-(-\mathbf q)+a_-(\mathbf q)a_+(-\mathbf q)\big)
\end{equation}
Translation symmetry exchanges $a_+$ and $a_-$, thus it guarantees that $\Pi_{++}(\mathbf q)=\Pi_{--}(\mathbf q)$.  A  non-zero $\Pi_{+-}(\mathbf q)$ leads to the different propagators between $a_1$ and $a_S$. In the following we calculate $\Pi_{+-}(\mathbf q)$ in perturbation theory.

To get a non-zero contribution to $\Pi_{+-}(\mathbf q)$, we need to include short range four-fermion interaction:
\begin{equation}
    L'_\Psi=V_c\sum_{a=\pm} (\Psi^\dagger_a \Psi_a)^2+ V \Psi^\dagger_+ \Psi_+ \Psi_-^\dagger \Psi_-
\end{equation}
In Appendix~\ref{appendix:photon_self_energy} we show that $\Pi_{+-}(\mathbf q)=0$ up to $O(V^2)$.    This suggests $r=0$ up to second of the interaction.  At the order $O(V^3)$, a non-zero $r$ may be generated, but we expect it to be quite small given that the four-fermion interactions are irrelevant.    The same result holds for contributions from the Higgs boson $\Phi$.  In summary, we have argued that the difference between the gauge couplings, $r$ in (\ref{defr}), is quite close to $0$.

\subsection{RG flow from $\epsilon$ expansion}

It is inspiring to control the calculation with $\epsilon$ expansion \cite{MMSS14}. We assume the action for the gauge fields is
\begin{equation}
   S_a= \frac{1}{2} \int \frac{d \omega d^2 q}{(2\pi)^3 } \left[ \left( \frac{1}{e_c^2}|q_y|^{1+\epsilon} +\kappa_0 \frac{|\omega|}{|q_y|} \right) |a_1(\omega,\mathbf q)|^2+ \left(\frac{1}{e_s^2}|q_y|^{1+\epsilon}+\kappa_0 \frac{|\omega|}{|q_y|} \right) |a_S(\omega,\mathbf q)|^2 \right]
\end{equation}

We should use $\epsilon=1$, but it is easier to do the calculation around $\epsilon=0$.  We define
\beq
\alpha_c=\frac{e_c^2\upsilon_F}{4\pi^2 \Lambda^\epsilon} \quad, \quad  \alpha_s=\frac{e_s^2\upsilon_F}{4\pi^2 \Lambda_y^\epsilon}\,, 
\eeq
where $\upsilon_F$ is the fermi velocity of $\Psi$ and $\Lambda_y$ is a cut off for $q_y$.

In Appendix~\ref{append:RG_flow}, we find the RG flow equations:
\begin{align}
    \frac{d\alpha_c}{d\ell}&=\frac{\epsilon}{2}\alpha_c-(\alpha_c+\alpha_s)\alpha_c \notag\\
    \frac{d\alpha_s}{d\ell}&=\frac{\epsilon}{2}\alpha_s-(\alpha_c+\alpha_s)\alpha_s 
\end{align}
There is a line of fixed points satisfying: $\alpha_c+\alpha_s={\epsilon}/{2}$ and $d (\alpha_c/\alpha_s)/d\ell = 0$. Because $\alpha_s(\ell=0)-\alpha_c(\ell=0) =r \alpha_c (\ell = 0)$ with $r \ll 1$, then at the fixed point, $\alpha_c\approx \alpha (1-{r}/{2})$ and  $\alpha_s \approx \alpha(1+{r}/{2})$, where $\alpha={\epsilon}/{4}$.
So we reach the important conclusion that in the interactions mediated by the two gauge fields have a nearly equal strength.

\subsection{Instability to pairing}

The leading contribution in BCS channel for ghost Fermi surface $\Psi$ is from exchange of one photon.  Generically, it is in the form
\begin{equation}
  S_{BCS}=\int d^2k_i d \omega_i \Psi^\dagger_a (k_1)\Psi^\dagger_b (-k_1)\Psi_d (-k_2)\Psi_c (k_2) \, V_{ab} \delta_{ac} \delta_{bd} \, F(k_1-k_2)
\end{equation}
where $F(q=k_1-k_2)$ arises from the propagator of the photons. The $\delta_{ac} \delta_{bd}$ factor is from the fact that the $U(1)_1$ and $U(1)_S$ interactions are diagonal in the $(\Psi_+,\Psi_-)$ basis.
For $a_1$, we have $V_{11}=1$, $V_{12}=1$.  For $a_S$, we have $V_{11}=1$ and $V_{12}=-1$.  This implies, as noted above, that $a_S$ mediates attractive interaction between $\Psi_+$ and $\Psi_-$, while $a_1$ mediates repulsive interaction between $\Psi_+$ and $\Psi_-$.  The final sign of the interaction between $\Psi_+$ and $\Psi_-$ depends on the competition between $a_1$ and $a_S$, and we seen in the subsections above that these interactions are nearly balanced.

We define dimensionless BCS interaction constant 
\begin{equation}
  \tilde V_m=\frac{k_F}{2\pi \upsilon_F} V_m
\end{equation}
where $m$ is the angular momentum for the corresponding pairing channel.
By integrating photon in the intermediate energy, we obtain \cite{MMSS14}
\begin{align} 
  \delta \tilde V_m&=\frac{k_F}{2\pi \upsilon_F} \upsilon_F^2 \int \frac{d\theta}{2\pi} \left(\frac{e^{-i m\theta}}{|k_F \theta|^{1+\epsilon}/e_{c}^{2}}-\frac{e^{-i m\theta}}{|k_F \theta|^{1+\epsilon}/e_{s}^{2}} \right)\notag\\
  &= \frac{\upsilon_F}{4\pi^2} 2 \int_{\Lambda e^{-\delta \ell/2}}^{\Lambda} dq_y \left(\frac{1}{|q_y|^{1+\epsilon}/e_{c}^{2}}-\frac{1}{|q_y|^{1+\epsilon}/e_{s}^{2}}\right)\notag\\
  &=(\alpha_c-\alpha_s)\delta \ell \label{BCSRG}
\end{align}

As shown in the previous subsection, we expect $\alpha_c\approx \alpha (1-{r}/{2})$ and  $\alpha_s \approx \alpha(1+{r}/{2})$, where $\alpha={\epsilon}/{4}$. Then $\alpha_c-\alpha_s=-r \alpha$.
The renormalization in (\ref{BCSRG}) should be combined with the usual flow of the BCS interaction \cite{MMSS14} to obtain the RG equation
\begin{equation}
  \frac{dV}{d\ell}=-r\alpha -V^2 \,. \label{RG2}
\end{equation}
The pairing stability then depends on the sign of $r$, which depends on non-universal details, such as the sign of $V$. Here we assume $0<r \ll 1$.  Then $V$ always flows to $-\infty$ regardless of the initial value.  For $0<V(\ell=0) \ll 1$, we find $V$ grows to $-\infty$ after $\ell^*= {\pi}/({2\sqrt{r\alpha}})$, suggesting a pairing scale $\Delta \sim \Lambda_\omega  e^{- {\pi}/({2\sqrt{r\alpha}})}$.  Given that $r \ll 1$, the pairing scale is exponentially suppressed.  There is another energy scale $E_{nFL}\sim \Lambda_\omega e^{-{1}/{\alpha}}$, below which we can observe non-Fermi-liquid behavior because of the gauge field fluctuations.   As $\alpha={\epsilon}/{4}={1}/{4}$, we expect $\Delta \ll E_{nFL}$ because $r\ll1$. This is opposite from the case of the Ising-nematic critical point \cite{MMSS14}. Consequently, we have obtained one of the important features of our theory: we expect a large energy window of non-Fermi-liquid behavior even though the critical point is covered by pairing.

\subsection{Phase diagram}
\begin{figure}[ht]
\centering
\includegraphics[width=0.8\textwidth]{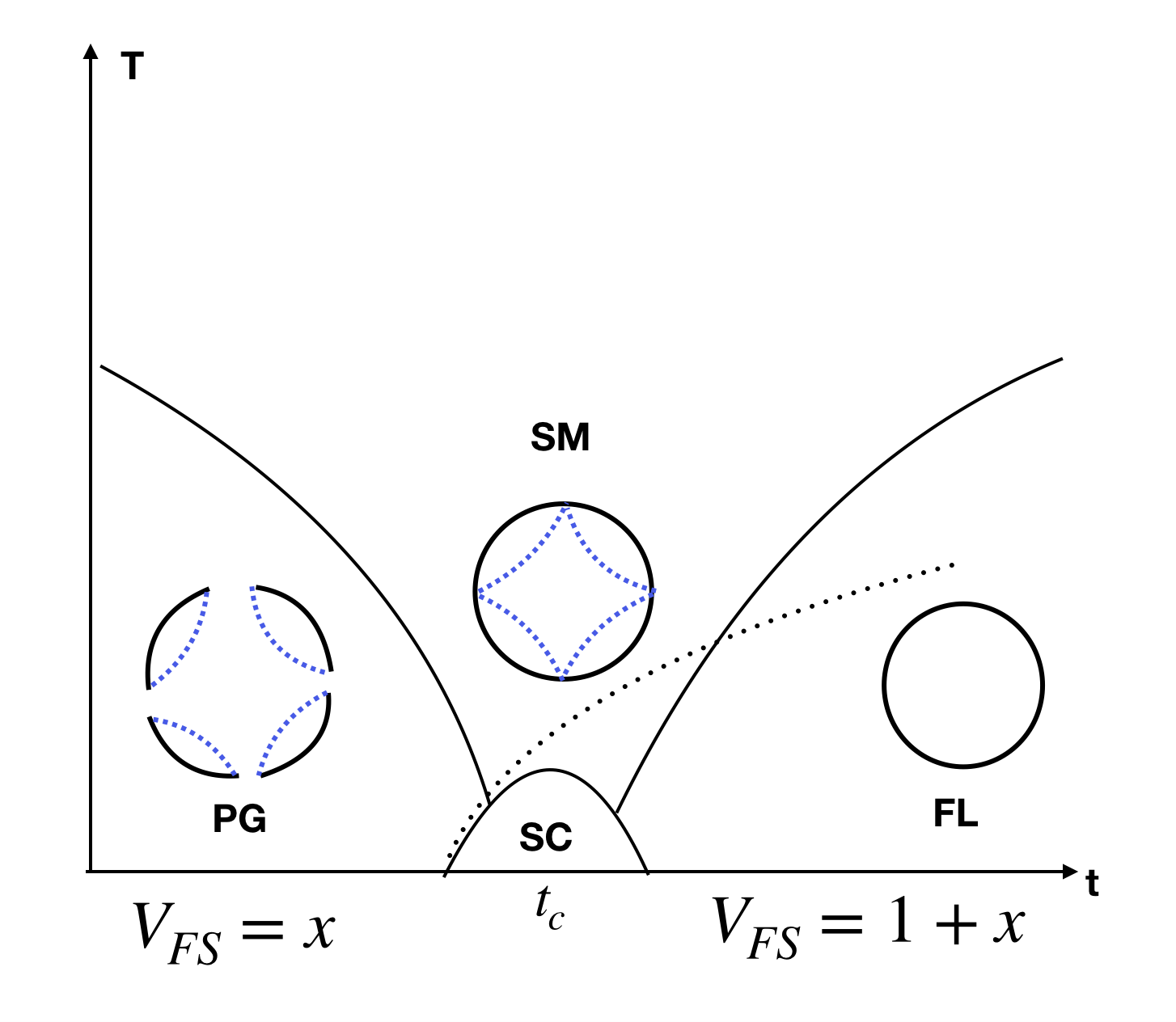}
\caption{A schematic phase diagram with $t$ a tuning parameter, such as doping, pressure or magnetic field, which effectively tunes the chemical potential $\mu_\Phi$ in our effective field theory. $T$ is the temperature.  $\langle \Phi \rangle \neq 0$ when $t<t_c$ and $\langle \Phi \rangle=0$ when $t>t_c$. $V_{FS}$ is the total Fermi surface volume and $x$ is the doping level.  The two crossover boundaries are the usual V-shaped quantum critical region for the critical boson $\Phi$.  The dashed line denotes the pairing scale of $\Psi$.
The blue dashed line denotes the Fermi surface from the ghost fermion $\Psi$, and the black line is the Fermi surface from physical fermion $C$.   }
\label{fig:phase_diagram}
\end{figure}

Based on the above analysis, we sketch a phase diagram in Fig.~\ref{fig:phase_diagram}. The critical point is driven by the condensation of $\Phi$.    At the critical point corresponding to the onset of $\langle \Phi \rangle \neq 0$, there can already be an exponentially small pairing $\langle \Psi \Psi \rangle \neq 0$ (in the case of the FL-FL* transition, the pairing instability is strong).  This means that the onset of the pairing $\langle \Psi \Psi \rangle \neq 0$ already happens in the $\langle \Phi \rangle \neq 0$ side.  When $\langle \Phi \rangle \neq 0$, the pairing of $\Psi$ will be transmitted to the pairing of $C$, and we obtain a superconductor at zero temperature.  When we enter into the side with $\langle \Phi \rangle =0$, $C$ can no longer inherit the pairing of the ghost fermion, because $C$ and $\Psi$ decouple now.  So the superconducting $T_c$ should decrease rapidly after $\langle \Phi \rangle=0$.  In this case, the pairing of $\Psi$ gaps out the ghost fermion, and then the remaining $U(1)_1 \times U(1)_S$ gauge fields just confine in $2+1$ dimensions, and we are left with only a physical Fermi surface.  (In $3+1$ dimension, the photon corresponding to $U(1)_1 \times U(1)_S$ gauge field can remain deconfined just after $\langle \Phi \rangle=0$, although it would be hard to detect this in experiments.) The pairing scale of $\Psi$ continues to increase into the Fermi liquid side because $U(1)_S$ gauge fluctuations become more like $SU(2)_S$
(see Appendix~\ref{append:RG_U2}) as $M_{1,2}$ in (\ref{HAF}) decrease.

If we start from the Fermi liquid side, there will be a ghost Fermi surface emerging around the critical region, which hybridizes with the physical Fermi surface and leads to partial Mott localization and a jump of carrier density.   The critical point is unstable to pairing at sufficiently low scale, but at finite temperature we expect the critical region to be governed by a stable DQCP.   The structure of the pseudogap phase depends on details. For the one orbital model, we will have small pockets formed by both $C$ and $\Psi$. In this case Fermi arcs are naturally expected in ARPES measurement because the ghost fermion $\Psi$ is invisible.   For the two-orbital Kondo model, we expect a small physical pocket mainly from $C$.

\section{Higgs boson fluctuations}
\label{sec:higgs_boson}

This section will further discuss the Higgs boson sector for the three DQCPs. 
We already presented the Lagrangians $L_\Phi$ in Section~\ref{sec:Higgsaction}.

The linear time derivative in the boson Lagrangians implies that the boson fluctuations are characterized by a dynamic critical exponent $z=2$, and consequently $d=2$ is upper-critical dimension. The influence of the couplings of the bosons to the ghost fermions and the gauge fields are quite similar to those discussed some time ago for the FL-FL* transition \cite{SVS04,Pepin1,Pepin2,Pepin3}: these couplings are unimportant for the leading critical behavior. The quartic interactions between the bosons are marginal in $d=2$, and we describe their one-loop RG equations below; we find that there are always regimes in which these interactions are marginally irrelevant.

The boson fluctuations can also contribute to the diamagnetic susceptibilities of the photons, as computed for the ghost fermions in Section~\ref{sec:diamagfermion}. However, for the linear time-derivative actions in Section~\ref{sec:Higgsaction}, this contribution is exactly zero when the bosons are not condensed; this is because the boson density vanishes in the scaling limit in the non-condensed phase \cite{ssbook}. We have to include irrelevant second order time-derivatives for a non-zero contribution. Even then, as in Section~\ref{sec:diamagfermion} and Appendix~\ref{appendix:photon_self_energy}, the low order diagrams will give an equal contribution to the two gauge fields, and a difference potentially appears only at third-order in the boson-boson quartic interactions.

\subsection{DQCP1}

Ignoring gauge fields, the boson theory for DQCP1 in (\ref{eq:Higgs_boson_DQCP1}) can be written as
\beq
L_\Phi = \Phi^\dagger_{a \alpha} \left[ \partial_\tau - \nabla^2 \right] \Phi_{a \alpha} + \gamma \, \Phi^\dagger_{a \alpha} \Phi^\dagger_{b \beta} \Phi_{b \beta} \Phi_{a \alpha} + \gamma_1 \, \Phi^\dagger_{a \alpha} \Phi^\dagger_{b \beta} \Phi_{a \beta} \Phi_{b \alpha}
\eeq
with $\gamma = \lambda - 3 \lambda_1$ and $\gamma_1 = 4 \lambda_1$.
The one-loop RG equations for the quartic boson-boson interactions are \cite{ssbook}
\bea
\frac{d \gamma}{d \ell} &=& - \frac{\gamma^2}{2 \pi} \nonumber \\
\frac{d \gamma_1 }{d \ell} &=& - \frac{\gamma_1^2}{2 \pi} - \frac{\gamma \gamma_1}{\pi}
\eea
Positive values of $\gamma, \gamma_1$ are therefore marginally irrelevant.

\subsection{DQCP2}

Similarly, the boson theory for DQCP2 in (\ref{eq:Higgs_boson:DQCP2}) can be written as
\beq
L_\Phi = \Phi^\dagger_{a \alpha} \left[ \partial_\tau - \nabla^2 \right] \Phi_{a \alpha} + \gamma \, \Phi^\dagger_{a \alpha} \Phi^\dagger_{a \beta} \Phi_{a \beta} \Phi_{a \alpha} + \gamma_1 \, \Phi^\dagger_{+ \alpha} \Phi^\dagger_{- \beta} \Phi_{+ \alpha} \Phi_{- \beta} + \gamma_2 \, \Phi^\dagger_{+ \alpha} \Phi^\dagger_{- \beta} \Phi_{+ \beta} \Phi_{- \alpha}
\eeq
with $\gamma = \lambda$, $\gamma_1 = \lambda_1 - \lambda_2$ and $\gamma_2 = 2 \lambda_2$.
The one-loop RG equations are
\bea
\frac{d \gamma}{d \ell} &=& - \frac{\gamma^2}{2 \pi} \nonumber \\
\frac{d \gamma_1}{d \ell} &=& - \frac{\gamma_1^2}{4 \pi} \nonumber \\
\frac{d \gamma_2}{d \ell} &=& - \frac{\gamma_2^2}{4 \pi} - \frac{\gamma_1 \gamma_2}{2 \pi}
\eea
Again, positive $\gamma,\gamma_1,\gamma_2$ are marginally attracted to zero coupling.

\subsection{DQCP3}
Finally, the boson theory for DQCP3 in (\ref{eq:Higgs_boson_DQCP3}) is already in a convenient form for RG analysis. Dropping the gauge fields, we have
\beq
L_\Phi = \varphi^\dagger_{a} \left[ \partial_\tau - \nabla^2 \right] \varphi_{a} + \lambda \left| \varphi_{a} \right|^4 + 
\lambda_1 \left| \varphi_{+} \right|^2 \left| \varphi_{-} \right|^2
\eeq
The one-loop RG equations are
\bea
\frac{d \lambda}{d \ell} &=& - \frac{\lambda^2}{2 \pi} \nonumber \\
\frac{d \lambda_1 }{d \ell} &=& - \frac{\lambda_1^2}{4 \pi} 
\eea
Now positive values of $\lambda, \lambda_1$ are marginally irrelevant.

\section{Non Fermi liquid behavior of the physical Fermi surface}
\label{sec:physical}

In Sections \ref{sec:critfermi} and \ref{sec:higgs_boson}, we focused on the ghost Fermi surface and the Higgs boson respectively.  Here we comment on the property of the physical Fermi surface formed by electron $C$, with an emphasis on possible non-Fermi liquid behavior. While the ghost Fermi surface dominates thermal properties, including the specific heat, as noted earlier, it is invisible to spin, charge, and photoemission probes.  The charge transport will be dominated by the physical Fermi surface and it is interesting to study whether the physical Fermi surface will become non-Fermi-liquid at the critical region.

We also emphasize a compelling feature of our theory. In our theory, the quasi particle scattering rate of $C$ will be equal to the transport scattering rate. This is because $\Psi$ is neutral, and the transport is dominated by $C$.  Although only a small momentum is transferred from $C$ to $\Psi$ in the process $C \rightarrow \Psi+\Phi^\dagger$, the current is lost into the ghost sector. Therefore, if there is non-Fermi-liquid behavior in quasi-particle scattering rate, there will also be a non-Fermi-liquid behavior in charge transport.

There are two main effects on the physical Fermi surface: ({\it i\/}) the Yukawa coupling 
\begin{equation}
    g\sum_{a=\pm} (C^\dagger \Phi_a) \Psi_a\,;
\end{equation} 
({\it ii\/}) the coupling to the order parameter:  
\begin{equation}
g_O\sum_i (C^\dagger C)(\Phi^\dagger \Phi)+g_s \sum_i (-1)^i (C^\dagger \vec \sigma C)\cdot (\Phi^\dagger \tau_z \vec \sigma \Phi)+g'_s \sum_i (C^\dagger \vec \sigma C)\cdot (\Phi^\dagger \vec \sigma \Phi)\,.    
\end{equation}
The coupling $g_O C^\dagger C O$ is irrelevant. To see this, we focus on a patch near $(k_x,k_y)=(k_f,0)$ around the Fermi surface. The dominant interaction is from $O(i\omega, \vec{q})$ with $\mathbf q$ in $y$ direction.  The action can be written as:
\begin{align}
S&= \int d\omega dk_x d k_y C^\dagger(-i\omega+\upsilon_Fk_x+\frac{k_y^2}{2m})C + \int dp_0 dp_x dp_y \Phi^\dagger(-ip_0+p_y^2)\Phi\notag\\
&~~~+g_O\int d\omega dk_x dk_y \int dp_0 d p_x dp_y \int dq_0 dq_x dq_y C^\dagger(k+q)C(k) \Phi^\dagger(p-q)\Phi(p)
\end{align}
where we have suppressed the index for $C$ and $\Phi$.  We have the scaling dimension $[k_y]=[p_y]=[q_y]=1,[k_x]=[p_x]=[q_x]=2, [\omega]=[p_0]=[q_0]=2, [\Psi]=-\frac{7}{2}, [\Phi]=-\frac{7}{2}$. Then it is easy to find $[g_O]=-1$.

The effect of the Yukawa coupling $g$ depends on the fermiology at the QCP.  If the Fermi surfaces of $C$ and $\Psi$ do not touch each other, this coupling is irrelevant. However, if the separation $\delta$ between the two Fermi surfaces is close to zero, there is a small energy scale $E_0$ above which the quasiparticle of $C$ can decay to a pair of $\Phi$ and $\Psi$,  similar to that described by Paul {\it et al.} \cite{Pepin1,Pepin2,Pepin3}.    For the Kondo lattice model, $\delta$ can be close to zero given that the Fermi surface going through the Mott transition is almost half filled.  For a single Fermi surface with density $1+p$, ${\delta}/{k_F}={p}/{2} \ll 1$ at the small $p$ if the Fermi surface is in a good cirular shape. However, if the Fermi surface has curvature, $C$ and $\Psi$ can be close to each other around several discrete hot spots (for example, around four anti-nodes in the square lattice for ansatz in Ref.~\onlinecite{Yahui1}).

\begin{figure}[ht]
\centering
\includegraphics[width=0.95\textwidth]{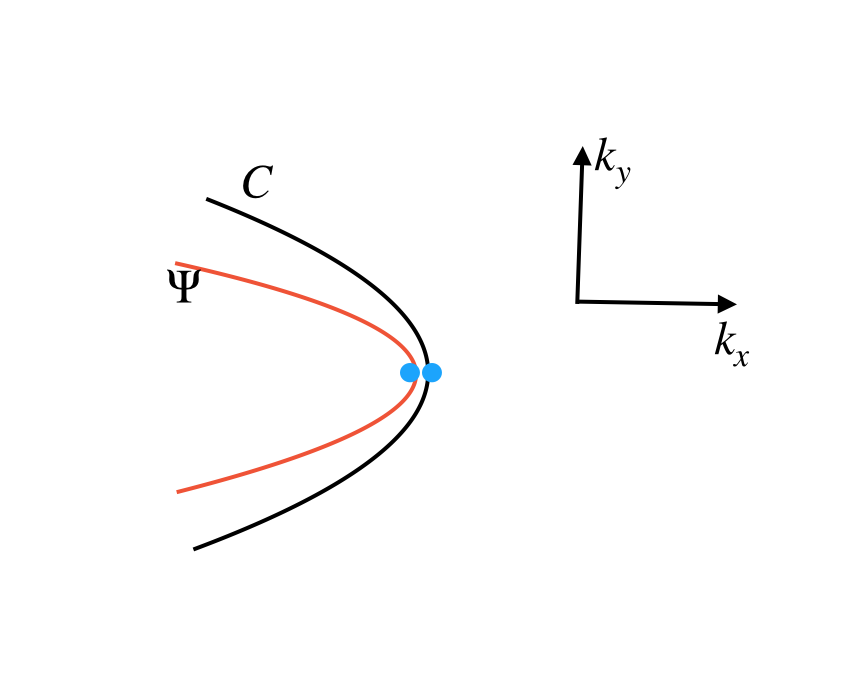}
\caption{An illustration of hot spots. $\delta$ is the distance between the two blue points. We assume $\delta \approx 0$. $C$ and $\Psi$ have opposite sign of dispersion. If $C$ forms an electron pocket, $\Psi$ forms a hole pocket.}
\label{fig:hot_spots}
\end{figure}

In the following we discuss possible non-Fermi liquid behavior in the context of heavy fermion systems.  We will focus on three dimension as many heavy fermion systems are three dimensional.  We will show that there is linear $T$ resistivity in the weak coupling regime. In the strong coupling regime where $g$ is large, we argue that the AF metal side will develop CDW order.

We assume $C$ and $\Psi$ have Fermi surfaces with spherical shapes. There should be two pockets for $C$ (see Fig.~\ref{fig:apres_AF_metal}(a)): one is a heavy pocket and is almost half-filled, and the other one is a light pocket and can have arbitrary filling.  In the following we will only consider the heavy pocket because the main role of the  $\Phi C^\dagger \Psi$ term is to Mott-localize this heavy pocket by hybridizing it with $\Psi$.  This heavy pocket of $C$ should have the opposite sign of dispersion as $\Psi$. For example we assume $C$ is an electron pocket while $\Psi$ is a hole pocket.  A small $\Phi$ thus can gap them out. 

We use the dispersion
\begin{align}
\xi_c(\mathbf k)&=\upsilon_c k \notag\\
\xi_\psi(\mathbf k)&=-\upsilon_\psi(k+\delta)
\end{align}
where $k=|\mathbf k|-k_F$ is defined relative to $k_F$ of C. Here we assume $M_1$ is very small at the QCP, so we can just use a spherical Fermi surface for $\Psi$.   We also have 
\begin{align}
\xi_c(\mathbf k+\frac{\mathbf q}{2})&=\upsilon_c (k+\frac{q}{2} \cos \theta) \notag\\
\xi_\psi(\mathbf k-\frac{\mathbf q}{2})&=-\upsilon_\psi (\delta+k-\frac{q}{2} \cos \theta)
\end{align}
where $\theta$ is the angle between $\mathbf k$ and $\mathbf q$.  $q=|\mathbf q|$ and we only keep terms up to $O(q)$, assuming $q$ is small.

Further analysis of the self-energy of the $C$ electron and boson $\Phi$ is describe in 
Appendix~\ref{app:marginal}.  First, $\Phi$ acquires a self-energy by decaying into a pair of $C$ and $\Psi$.  In the region $|q_0| \ll \sqrt{\upsilon_c \upsilon_\psi} \delta$ and $q \ll \delta$, we obtain:
\begin{equation}
  \Sigma_\Phi(iq_0,\mathbf q) \approx \frac{g^2k_F^2}{2\pi^2(\upsilon_c+\upsilon_\psi)} \left[ -i \left(\frac{\upsilon_c-\upsilon_\psi}{\upsilon_c}\right) \frac{q_0}{\upsilon_\psi \delta}- \frac{1}{3} \frac{q^2}{\delta^2} \right]
\end{equation}
where we keep only $O(q^2)$ and $O(q_0)$ terms. 
We have $G^{-1}_{\Phi}(0,\mathbf q)=\frac{1}{2m_\Phi}(1-\frac{2m_\Phi}{3 \delta^2}\alpha)q^2$ where $\alpha= \frac{g^2 k_F^2}{2\pi^2(\upsilon_c+\upsilon_\psi)}$.  If $\frac{2m_\Phi}{3 \delta^2} \alpha>1$, then the $\Phi$ will have minimum at non-zero $\mathbf q$.   In the following we discuss the weak coupling region $\frac{2m_\Phi}{3 \delta^2} \alpha<1$  and the strong coupling region  $\frac{2m_\Phi}{3 \delta^2} \alpha>1$ separately.

\subsection{Marginal Fermi liquid at weak coupling}

In the following we consider the scattering rate of the process $C(\mathbf k) \rightarrow \Phi(\mathbf q)+ \Psi(\mathbf k')$.  This can be done by using the Fermi's golden rule.  We need a momentum $q\sim \delta$ for the boson $\Phi$ to compensate the momentum mismatch between $C$ and $\Psi$, so the following result only applies above a small energy scale $E_0=\frac{\delta^2}{2m_\Phi}$.   In Appendix.~\ref{app:marginal}, we find that the scattering rate for $C$ with energy $\xi_C(\mathbf k)$ is

\begin{equation}
    \Sigma''_C(\mathbf k)=f \xi_c(\mathbf k)
\end{equation}
where
\begin{equation}
  f=\frac{\alpha  2\tilde m_\Phi \pi}{6k_F^2}=\frac{\pi}{2} \frac{\delta^2}{k_F^2} \frac{\tilde \alpha}{1-\tilde \alpha}
\end{equation}
with $\tilde \alpha=\frac{2m_\Phi}{3 \delta^2} \alpha$.

We find that the scattering rate $(\frac{1}{\tau})_{qp}=f \xi_c(\mathbf k)$ when $\chi_c(\mathbf k)>E_0$ and it is natural to expect $(\frac{1}{\tau})_{qp}=f k_B T$ at finite temperature $T>E_0$. 

Linear $T$ resistivity has been observed in many heavy fermion systems around the critical region.  Our theory could be an explanation.   Let us also briefly comment on two dimension.  In two dimension, the ghost Fermion $\Psi$ will acquires a self energy $\Sigma_\psi(\omega)\sim |\omega|^{\frac{2}{3}} \sign(\omega)$ from coupling to the gauge fields.  This makes the analysis of the $g \Phi C^\dagger \Psi$ term challenging.  Besides, in the context of cuprate, the Fermi surface shape of $C$ and $\Psi$ are likely not perfectly circular and they may have small $\delta$ only around hot spots (around the anti-node region). This situation needs an independent analysis.  We thus leave a detailed study to future work.

\subsection{Strong coupling regime: Instability of ``FFLO'' order}

\begin{figure}[ht]
    \centering
    \includegraphics[width=0.9\textwidth]{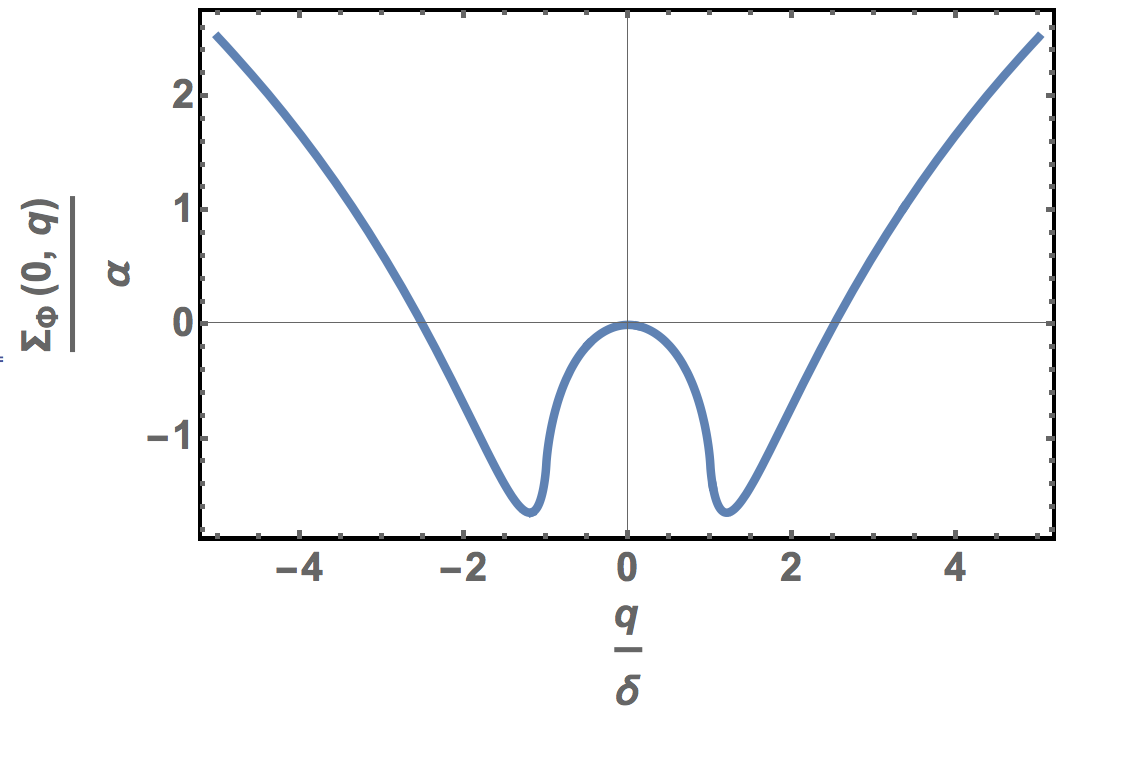}
    \caption{$\Sigma_\Phi (0,\mathbf q)$ has a minimum around $|\mathbf q|\approx \delta$.}
    \label{fig:FFLO_instability}
\end{figure}

As computed in Appendix~\ref{app:hm}, in the strong coupling regime (with $\frac{2m_\Phi}{3 \delta^2} \alpha>1$ or $\tilde \alpha>1$), $G^{-1}_{\Phi}(0,\mathbf q)=\frac{1}{2m_\Phi}(1-\frac{2m_\Phi}{3 \delta^2}\alpha)q^2$ at small $q$ has a negative coefficient in front of $q^2$ term.   Actually, when $\alpha$ is large, $G^{-1}_{\Phi}(0,\mathbf q)$ is dominated by $\Sigma_\Phi(0,\mathbf q)$ and has minimum at $|\mathbf q| \approx \delta$ (see Fig.~\ref{fig:FFLO_instability}), suggesting an instability of condensing $\Phi$ at non-zero momentum $Q\approx \delta$. 

$\Phi C^\dagger \Psi$ is analogous to a superconductor term because $C$ and $\Psi$ have opposite sign of dispersion.   When there is a Fermi surface mismatch with $\delta \neq 0$, the situation is similar to that of pairing with spin imbalance under a Zeeman field.  The instability at $Q \approx \delta$ is thus an analog of the Fulde-Ferrell-Larkin-Ovchinnikov (FFLO)
superconductor.   In our case, a condensation with $\langle \Phi \rangle=|\Phi_0| e^{i \mathbf{Q} \cdot \mathbf r}$ corresponds to a CDW order parameter coexisting with the AF metal. 
 
The occurrence of this FFLO type of instability needs strong coupling $\tilde \alpha$. A similar instability also exists in two dimension.   Interestingly CDW order has been found in underdoped cuprate (especially at high magnetic field and low temperature).  A pseudogap metal  with CDW order can be obtained by modifying our description of FL* phase with  $\Phi$ condensing at a non-zero momentum $Q \sim \delta$.  We will explore this possibility in future.

Similar ``FFLO'' physics has been discussed in  Ref.~\onlinecite{Pepin1}. But in this previous study the boson is an exciton formed by two physical pockets.  In contrast, the boson in our theory is an exciton between physical pocket and a ghost Fermi surface.   A small $\delta$ in Ref.~\onlinecite{Pepin1} needs the Fermi surfaces of the two physical pockets to almost coincide, which needs certain level of fine-tuning. For example, the heavy pocket is usually half-filled while the light pocket can have generic filling, suggesting that a small $\delta$ is not easy to satisfy.    In our theory, the ghost Fermi surface is guaranteed to be half-filled. Besides we interpret $\Psi$ as correlation hole of $C$, thus its Fermi surface is likely to follow that of $C$ and a small $\delta$ is more natural in our theory.

\section{Discussion}
\label{sec:conc}

In this paper, and in our earlier work \cite{Yahui1}, we have shown that ancilla qubits are a powerful tool in resolving many long-standing issues in the theory of quantum phase transitions of correlated metals. When studying phase transitions between states with distinct Fermi surfaces, and more so in cases where the Fermi surfaces carry distinct gauge charges, past approaches invariably made a choice of one set of Fermi surfaces about which to analyze fluctuations; this usually led to difficulties in describing `the order side' of the phase transition, where the emergence of a new set of fermions was required from non-perturbative effects. The ancilla approach is more democratic, and allows one to treat both sets of Fermi surfaces within the same framework, and within perturbation theory. And subtle constraints on the relationship between Fermi surface volume \cite{SVS04,Paramekanti_2004} and the bulk topological order become much easier to implement on both sides of the transition.

We have shown here that the ancilla approach allows to obtain a specific theory for a long-standing open problem in the study of the heavy-fermion compounds and the cuprate superconductors: a transition with simultaneous Kondo breakdown and onset of magnetic order. We obtain specific variational wavefunctions which can be extended across the transition. And we have also a presented a critical theory (DQCP2) for the quantum fluctuations near the transition.

The main physical prediction of our approach is appearance of ``ghost'' Fermi surfaces near the quantum phase transition. At first sight, this might appear to be an artifact of the fact that we have introduced additional ancilla degrees of freedom in the hidden layers. However, our careful treatment of the gauge fluctuations shows that this is not the case: the ghost fermions are excitations of the physical Hamiltonian, and should be detectable in experiments. On the issue of `overcounting' degrees of freedom, we make the following remarks:
\begin{itemize}
    \item Fock space is exponentially large, and there is plenty of space for new quasiparticle species.
    \item A solvable model without quasiparticles is the Sachdev-Ye-Kitaev model. This has a non-zero entropy in the limit of zero temperature \cite{GPS01}, implying of order $e^N$ states at energies above the ground state of order $1/N$ ($N$ is the system size).
Adding another Fermi surface implies exponentially fewer low energy states.
    \item Even in non-metallic states without quasiparticles {\it e.g.\/} strongly coupled conformal field theories in 2+1 dimensions, there an infinite number of primary operators (which are analogs of quasiparticles).
\end{itemize}

Experimental detection of the ghost Fermi surfaces likely requires sensitive thermal probes which can account for all the low energy degrees of freedom. Measurements of the thermal conductivities and the specific heat are possibilities. Furthermore, as the ghost Fermi surfaces do not carry spin or charge, observation of Fermi surfaces in thermal probes, along with their absence in spin or charge measurements, would constitute a unique signature.
We expect a great enhancement of density of states around the critical region, and this can be tested by measurement of $C/T$, where $C$ is the specific heat.  In many other theories, enhancement of $C/T$ is also predicted because of the increasing of the mass of physical Fermi surface.   In our theory, there is an additional large contribution from just the ghost fermion.   In experiments, one can fit the mass of the physical Fermi surface from another independent probe (such as quantum oscillation) and therefore isolate the contribution from the physical Fermi surface.  An additional large contribution to $C/T$ beyond that from the physical Fermi surface is a falsifiable prediction of our theory.

In the following subsection we will present an intuitive interpretation of our results, noting the key role played by the ancilla. 
This will be followed in Section~\ref{sec:dcp} by a summary of the deconfined criticalities of Fig.~\ref{fig:global_phase_diagram}.

\subsection{Physical meaning of ancilla qubits and variational wavefunctions}

The limit $J_\perp \rightarrow +\infty$ in Fig.~\ref{fig:layers} leads to the constraint in (\ref{Sconstraints}).  In this limit, naively, the ancilla qubits decouple from the physical layer, but actually we can still write down a variational wavefunction without decoupling while respecting the constraint in (\ref{Sconstraints}) exactly \cite{Yahui1}. 

\begin{figure}
    \centering
    \includegraphics[width=0.75\textwidth]{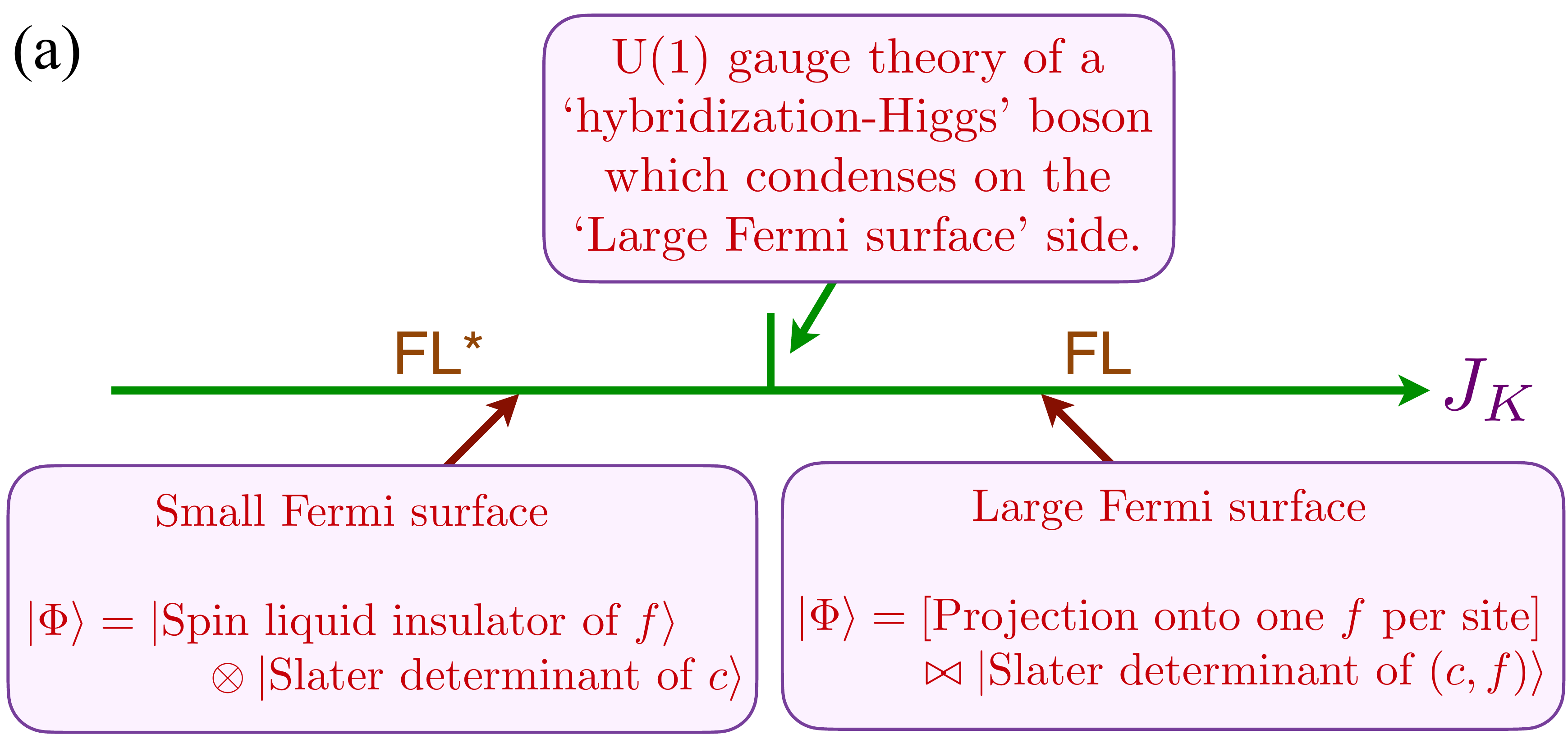}\\~\\
        \includegraphics[width=0.75\textwidth]{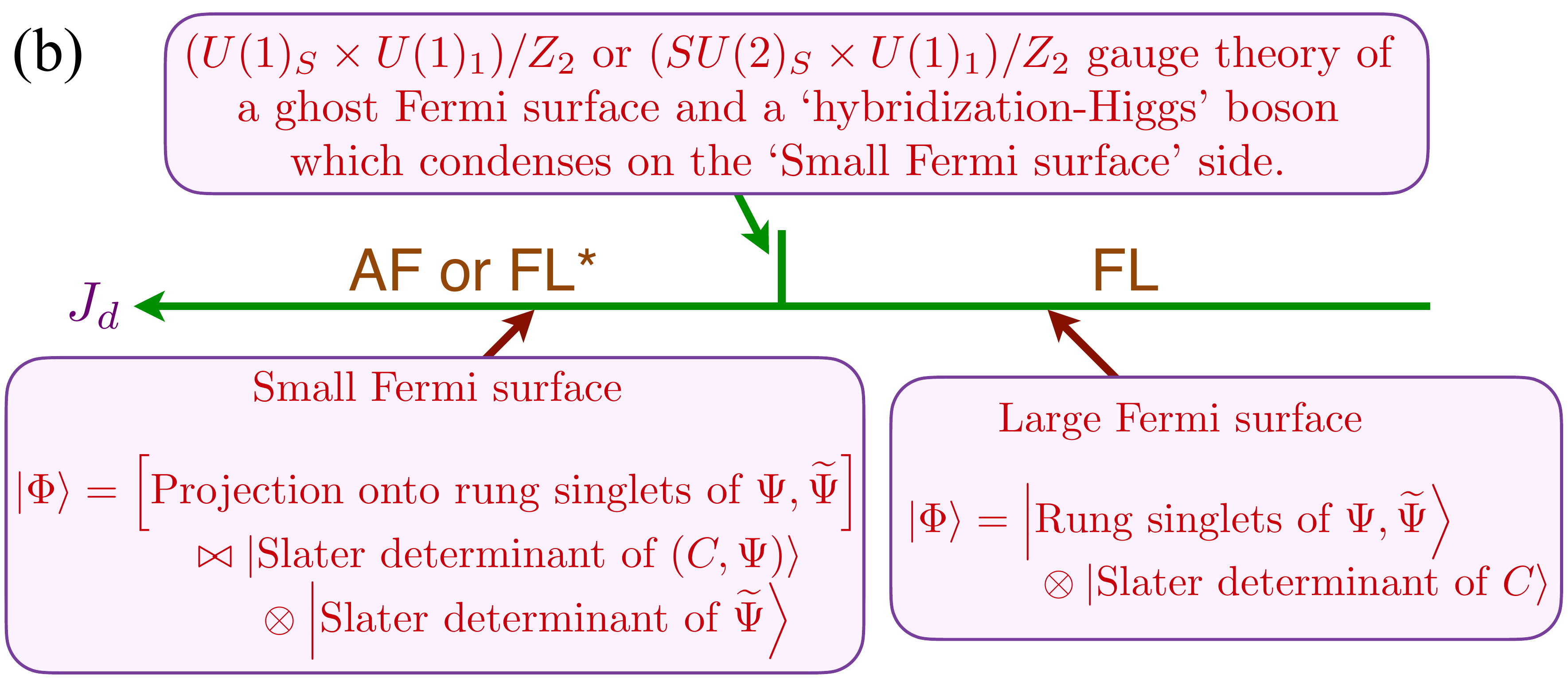}
    \caption{Comparison of variational wavefunctions $\left| \Phi \right\rangle$ for the previous theory \cite{SVS04} of the FL*-FL transition with the ancilla approach. (a) Previous theory of a Kondo lattice model of $f$ spins coupled via a Kondo exchange $J_K$ to $c$ conduction electrons. (b) Ancilla model of Fig.~\ref{fig:layers}. Note that the Higgs phases are on opposite sides in the two cases. }    \label{fig:wavefunction}
\end{figure}

A class of variational wavefunctions for the pseudogap metal is of the form
\begin{equation}
    \ket{\Phi}= \sum_{a}  \braket{a,s| \text{Slater}[C,\Psi] \, \text{Slater}[\widetilde \Psi]} \ket{a}
    \label{eq:variational_wavefunction}
\end{equation}
where $a$ is summed over the many-body basis of the physical Hilbert space. $\ket{s}=\prod_i \frac{1}{\sqrt{2}}\epsilon_{ab} \Psi^\dagger_{i;a} \widetilde \Psi^\dagger_{i;b} \ket{0}$ is a trivial product state for the ancilla qubits.  $\text{Slater}[C,\Psi]$ is a Slater determinant corresponding to a mean field ansatz for $C$ and $\Psi$ with possible coupling $g C^\dagger \Psi$. $\text{Slater}[\widetilde \Psi]$ is another Slater determinant for $\widetilde \Psi$.  
See Fig.~\ref{fig:wavefunction} which contains a comparison of such wavefunctions with earlier work on Kondo lattice models.

The state $\ket{\Phi}$ is a physical state purely in the physical Hilbert space. If the coupling $g \neq 0$, the ancilla qubits do not disappear and they actually influence the physical wavefunction.  So they clearly have physical meaning.  When $g=0$, the wavefunction in (\ref{eq:variational_wavefunction}) factorizes, and a conventional Fermi liquid is obtained after projecting the ancilla qubits to the trivial product state $\ket{s}$.  When $g\neq 0$, $\Psi$ and $C$ hybridize, and part of $C$  gets ``Mott'' localized and only a small Fermi surface is itinerant. Consequently, the $\Psi$ can be viewed as correlation holes, which are responsible for the partial Mott localization \cite{Yahui1}.  We believe $\Psi$ field corresponds to many-body collective excitation in the physical Hilbert space, which is difficult to capture using conventional methods.  In our framework, we use gauge theory to include these possibly non-local collective excitations by introducing them as auxiliary degrees of freedom. After that, we can work in an enlarged Hilbert space where these collective excitations are viewed as elementary particles, and can be treated in a simple mean field theory. The physical Hilbert space can be recovered by projecting the auxiliary degree freedom to form trivial product state $\ket{s}$. The constraint can be equivalently implemented by introducing gauge fields, as we have discussed in the body of the paper. 
With the gauge constraint, $\Psi$ and $\widetilde \Psi$ are ghost fermions which carry neither spin nor charge.  A coupling $g\neq 0$ will Higgs the gauge field, and after that $\Psi$ can be identified as the physical electron and $\widetilde \Psi$ can be identified as a neutral spinon.  Because the density of $\Psi$ is unity, the $g$ coupling successfully describes a partial Mott localization of one electron per site from the large Fermi surface, while $\widetilde \Psi$ represents the local moment of the localized electron.  

In our theory, there can be a deconfined ghost Fermi surface in addition to the physical Fermi surface in  the critical region, leading to larger density of states. In the following we argue that additional degree of freedom is an intrinsic feature of the phase, not an artifact of our parton theory.  We can get inspiration from  a similar parton theory which has been developed to describe the composite Fermi liquid (CFL) for boson at $\nu=1$ \cite{PH98,read1998lowest,dong2020non}, where an auxiliary fermion is introduced to represent the correlation hole (or vortex). A gauge constraint is needed to project the state of the auxiliary fermion to recover the physical Hilbert space. In this theory, the number of single particle states is enlarged to $O(N_{\Phi}^2)$ compared to $O(N_\Phi)$, where $N_{\Phi}$ is the number of magnetic flux.  The additional degree of freedom arises from the inclusion of the correlation hole. It is now widely believed that the correlation hole (or vortex) plays an essential role in the CFL physics. Traditionally, the correlation hole is introduced through flux attachment. In contrast, in the theory of Refs.~\onlinecite{PH98,read1998lowest,dong2020non}, the correlation hole is included explicitly as an auxiliary fermion. After the gauge constraint is fixed, this theory can be shown to be equivalent to the conventional Halperin-Lee-Read theory based on flux attachment in the level of both the variational wavefunction \cite{read1998lowest} and the low energy field theory \cite{dong2020non}.  In CFL phase, the correlation hole is a real object and  auxiliary fermion  is a useful trick to represent it.   In the similar spirit, the additional fields $\Psi, \widetilde \Psi$ in our theory are also introduced to represent intrinsic many-body collective excitations. At the deconfined critical points, summarized in Fig.~\ref{fig:global_phase_diagram} and Section~\ref{sec:dcp}, there are emergent ghost Fermi surfaces and we believe they correspond to intrinsic non-local excitations responsible for the partial Mott localization. Unlike the quantum Hall systems, there is no other technique like flux-attachment to introduce these excitations, and the framework based on auxiliary degrees of freedom is likely the easiest way to include them in the low energy theory.

The detailed ansatz of these collective excitations can be determined numerically from optimizing the physical Hamiltonian using the variational wavefunction in (\ref{eq:variational_wavefunction}).  In this paper, we have studied critical points tuned by the mass of a Higgs boson corresponding to $C^\dagger \Psi$.  The correspondence of this Higgs mass to the microscopic model is not clear in our theory, but can in principle be determined numerically based on the variational wavefunction. We have focused on the universal properties here, and leave the energetics to future numerical studies.  

\subsection{Deconfined criticality}
\label{sec:dcp}

We have presented here a set of deconfined critical theories for the phase transitions in Fig.~\ref{fig:global_phase_diagram}, labeled DQCP1, DQCP2, DQCP3. The matter fields of these theories are:
\begin{itemize}
    \item Higgs fields, $\Phi$, carrying fundamental charges of emergent gauge fields, along with unit physical electromagnetic charge, $S=1/2$ under global spin rotations, and transformations under lattice and time-reversal symmetries.
    \item Ghost fermions, $\Psi$, which carry neither spin nor charge, and so are detectable only in energy probes.
    \item Fermi surface of the underlying electrons, $C$, which are not fractionalized in any stage of the theory.
\end{itemize}
The gauge sector had $(SU(2)_S \times U(1)_1)/Z_2$ gauge fields for DQCP1, and $(U(1)_S \times U(1)_1)/Z_2$ gauge fields for DQCP2 and DQCP3. The `$S$' gauge fields mediate an attractive interaction between the ghost fermions in the even parity channel, while the `$1$' gauge fields mediate a nearly equal repulsive interaction. We described the interplay between these interactions using the methods of Ref.~\onlinecite{MMSS14} in Section~\ref{sec:critfermi}, and showed that it was possible to find conditions under which the critical ghost Fermi surface could be stable to pairing above an exponentially small energy scale.

We contrast to other recent works on optimal doping criticality \cite{Chowdhury:2014efa,Sachdev:2018ddg,SSST19,SPSS20,SSS20}, which had only a $SU(2)_S$ gauge field, and Higgs fields that were neutral under physical electromagnetism and spin. There are no ghost fermions in these theories, but the cases with gauge-charged fermions carrying electromagnetic charge (`chargons') \cite{Chowdhury:2014efa,Sachdev:2018ddg} do have a pairing instability to superconductivity.

We noted earlier the recent work \cite{DQCPSG} on deconfined criticality in the optimally doped cuprates in models with disorder. Here, the overall picture of the phase diagram is similar to that of the present paper, except that the antiferromagnetic order is replaced by the spin glass order. In the limit of large spatial dimension, this work leads to a robust mechanism for linear in temperature resistivity \cite{Guo:2020aog}.

We also discussed some observable consequences of our DQCPs. DQCP1 and DQCP2 allow for a jump in the size of the Fermi surfaces, and a correspondingly discontinuous Hall effect. Photoemission experiments detect only  the $C$ Fermi surfaces, and these could have marginal Fermi liquid-like spectra, as discussed in Section~\ref{sec:physical}. We have already discussed thermal detection of the ghost fermions earlier in this section.
In the presence of disorder, the AF order near the DQCP could be replaced by glassy magnetic order.
These are all phenomenologically very attractive features \cite{Badoux16,Michon18,Julien19,Dessau19,Ramshaw2020}.
We speculate that the ghost fermions also play a significant role in the anomalous thermal Hall effect observed recently \cite{grissonnanche2019giant,grissonnanche2020phonons}.

\section*{Acknowledgements}

We are very grateful to G.~Vignale for sharing details of the computation in Ref.~\onlinecite{Vignale88} which was useful for the analysis in Appendix~\ref{appendix:photon_self_energy}.
We thank P.~Coleman, T.~Senthil, and G.~Vignale for useful discussions.
This research was supported by the National Science Foundation under Grant No.~DMR-2002850. This work was also supported by the Simons Collaboration on Ultra-Quantum Matter, which is a grant from the Simons Foundation (651440, S.S.).

\appendix

\section{Photon self-energy for $U(1)\times U(1)$ theory \label{appendix:photon_self_energy}}

For simplicity, let us organize the terms in the effective action for the gauge field as:
\begin{equation}
    L'_{\Phi}[a]=  \Pi_+(\mathbf q) |a_+(\mathbf q)|^2 +\Pi_-(\mathbf q)|a_-(\mathbf q)|^2+\Pi_{+-}(\mathbf q) \big( a_+(\mathbf q) a_-(-\mathbf q)+a_-(\mathbf q)a_+(-\mathbf q) \big)
\end{equation}
where the gauge fields $a_{\pm}$ were introduced above (\ref{LPsi23}); we view $a_{a}$ with $a=\pm$ as a probe gauge field coupling to $\Phi_a$ and $\Psi_a$.  Because the translation symmetry $T_x$ exchanges the two flavors, $\Pi_+(\mathbf q)=\Pi_-(\mathbf q)$.

From the linear relation $a_+=a_1+a_S$ and $a_-=a_1-a_S$, we can rewrite the action as

\begin{equation}
    L'_{\Phi}[a]=  \Pi_c(\mathbf q) |a_1(\mathbf q)|^2+ \Pi_s(q) |a_S(\mathbf q)|^2
\end{equation}
where
\begin{align}
    \Pi_c(\mathbf q)&= \Pi_+(\mathbf q)+\Pi_-(\mathbf q)+2 \Pi_{+-}(\mathbf q)\notag\\
    \Pi_s(\mathbf q)&=\Pi_+(\mathbf q)+\Pi_-(\mathbf q)-2 \Pi_{+-}(\mathbf q)
\end{align}

In the leading order, we find $\Pi_c(\mathbf q)=\Pi_s (\mathbf q)=\chi_f=\frac{1}{12\pi m_{\Psi}}$. Then we show that there is no correction for this result up to three loops upon including local interactions for the fermions and Higgs bosons.

\subsection{Leading order}

At leading order,  we consider one-loop bubble from fermion $\Psi$.  We only have $\Pi^{aa}_{xx}(\mathbf q)=P_{aa} q_y^2$. The one-loop bubble is

\begin{equation}
    \Pi^{aa}(q)= \sum_p   G_0(p-\frac{q}{2})G_0(p+\frac{q}{2}) \frac{p_x^2}{m^2}
\end{equation}
where,
\begin{equation}
    G_0(p)=\frac{1}{-i p_0+\xi(p)}
\end{equation}
with $\xi(p)=\frac{p^2}{2m_\psi}$.
We define
\begin{equation}
    R(p,q)=G_0(p-\frac{q}{2})G_0(p+\frac{q}{2})
\end{equation}
and we expand it around $q=0$:
\begin{equation}
    R(p,q)=R_0(p)+R_2(p)q^2
\end{equation}
Then we have
\begin{equation}
    P= \sum_p R_2(p) \frac{p_x^2}{m_\psi^2}
\end{equation}
From the expansion $G_0(p+q)=G_0(p)-\frac{\partial G_0(p)}{\partial \mu} (\xi(p+q)-\xi(p))+\frac{1}{2} \frac{\partial^2 G_0(p)}{\partial \mu^2} (\xi(p+q)-\xi(p))^2$ and $\xi(p+q)-\xi(p)=\frac{p_yq_y}{m_\psi}+\frac{q_y^2}{2m_\psi}$, we obtain
\begin{align}
    R_2(p)&=-\frac{1}{4m_\psi} G_0(p) \frac{\partial G_0(p)}{\partial \mu}+\frac{p_y^2}{4 m_\psi^2} \left(G_0(p)\frac{\partial^2 G_0(p)}{\partial \mu^2}-(\frac{\partial G_0(p)}{\partial \mu})^2 \right) \notag\\
    &=- \frac{1}{8 m_\psi} \frac{\partial^2 G_0(p)}{\partial \mu^2}+ \frac{p_y^2}{24m_\psi^2} \frac{\partial^3 G_0(p)}{\partial \mu^3}
\end{align}
Next we can first integrate $p_0$ and get $\sum_{p_0} G(p)=- \theta(\mu-\frac{p^2}{2m_\psi})$.  Finally, we can integrate $d^2 p=p dp d\theta= m_\psi d\epsilon d\theta $ to obtain
\begin{align}
    P&= \frac{m_\psi}{4 \pi^2} \int d\epsilon  d\theta  \cos^2 \theta 2\epsilon [-\frac{1}{8m_\psi} \delta'(\epsilon-\mu)-\sin^2 \theta \cos^2\theta \frac{\epsilon^2}{6 m_\psi} \delta''(\epsilon-\mu)]=\frac{1}{24 \pi m_\psi}
\end{align}
Finally, we find at this order that $\Pi_c(\mathbf q)=\Pi_s(\mathbf q)=\Pi_+(\mathbf q)+\Pi_-(\mathbf q)=2P q_y^2=\chi_f q_y^2$ with $\chi_f=\frac{1}{12 \pi m_\psi}$.

\subsection{Ward Identities}

Higher order computations of the diamagnetic susceptibilities are simplified by consideration of Ward identities.
There are two conserved quantities: $J^0_c(x)= \sum_a\Psi^\dagger_a(x) \Psi_a(x)$ and $J^0_s(x)=\sum_a \tau^z_{aa} \Psi^\dagger_a(x)\Psi_a(x)$.   For each charge, there is a continuty equation:

\begin{equation}
    \partial_\mu J^\mu_\alpha =0
\end{equation}
where $\alpha=c,s$ denotes the gauge field.
We also define vertex functions:
\begin{equation}
    G_a(p-\frac{q}{2})G_a(p+\frac{q}{2})\Lambda^{\mu}_{\alpha;a}(p;q)= \int d^{d+1}x \int d^{d+1}y \int d^{d+1}z e^{-i q x}e^{-i(p-\frac{q}{2}) y} e^{i (p+\frac{q}{2})z}\langle T J^{\mu}_\alpha(x) \Psi_a(y) \Psi^\dagger_a(z) \rangle
\end{equation}
where $T$ is the time ordering operator. In the above, $x,y,z$ are space-time coordinates. Similarly $p=(p_0,\vec{p})$ also contains the frequency.

We can apply $\partial_{x_\mu}$ to the above equation and use the continuity equation $\partial_\mu J^\mu_\alpha =0$.  There will be an additional contribution from derivative to the time ordering operator:
\begin{align}
    \partial_{x_0} \langle T J^{\mu}_\alpha(x) \Psi_a(y) \Psi^\dagger_a(z) \rangle&= \langle T \partial_0 J^0_\alpha(x) \Psi_a(y)\Psi^\dagger_z(z) \rangle+ \delta^{(d+1)}(x-y) \langle T [J^0_\alpha(x),\Psi_a(x)] \Psi^\dagger_a(z) \rangle \notag\\
    &~~~~+ \delta^{(d+1)}(x-z) \langle T  \Psi_a(y) [J^0_\alpha(x),\Psi^\dagger_z(x)] \rangle \notag\\
    &=\langle T \partial_0 J^0_\alpha(x) \Psi_a(y)\Psi^\dagger_z(z) \rangle-\delta^{(d+1)}(x-y)  \tau^{\alpha}_{a} \langle T \Psi_a(x) \Psi^\dagger_a(z) \rangle \notag\\
    &~~~~~~+\delta^{(d+1)}(x-z)\tau^{\alpha}_a \langle T \Psi_a(y) \Psi^\dagger_a(z) \rangle
\end{align}
where $\tau^c_a=\tau^0_{aa}$ and $\tau^s_a=\tau^z_{aa}$. We have used identities:
\begin{align}
[\Psi^\dagger_a(x) \Psi_a(x), \Psi_b(y)]&=-\delta_{ab} \Psi_b(x) \delta^{(d+1)}(x-y) \notag\\
[\Psi^\dagger_a(x) \Psi_a(x), \Psi_b^\dagger(y)]&=\delta_{ab} \Psi_b^\dagger(x) \delta^{(d+1)}(x-y) \notag\\
\end{align}
Finally we get
\begin{equation}
    q_\mu \Lambda^{\mu}_{\alpha;a}(p;q)=\tau^\alpha_a ( G^{-1}_a(p+\frac{q}{2})-G^{-1}_a(p-\frac{q}{2}))
\end{equation}
We focus on the case with $q_0=0$. By taking derivative to $q_\mu$ in both sides, we can derive that
\begin{equation}
    \Lambda^{\mu}_{\alpha;a}(p,q\rightarrow 0)=\tau^\alpha_a  \partial_{p_\mu}  G^{-1}_a(p)
    \label{eq:Ward_Identity}
\end{equation}
The same Ward Identity holds for the vertex between gauge field and Higgs boson $\Phi$.

\subsection{General result}

We try to derive a general form of the self energies of the photons, following the analysis in Refs.~\onlinecite{Vignale88}. We will illustrate the derivation for fermion $\Psi$, but the same result holds for boson $\Phi$.   We consider $\mathbf a$ in $x$ direction and use the Coulomb gauge $\mathbf q \cdot \mathbf a=0$. So  $q_x=0$ and we only need to consider momentum $(q_0,q_x,q_y)=(0,0,q)$. We can then use the abbreviation $\Pi_{\alpha}(q)=\Pi^{\alpha \alpha}_{xx}(\mathbf q)$, where $\alpha=c,s$ denotes the gauge field $a_1$ and $a_S$.

We have the following general relation
\begin{equation}
    \Pi_\alpha(q)=\sum_p \sum_{a=\pm} \lambda^{\alpha}_{a}(p) G_a(p+\frac{q}{2}) G_a(p-\frac{q}{2}) \Lambda^{\alpha}_{a}(p)
\end{equation}
where $\Lambda_{a}^{\alpha}(p,q)=\Lambda^{x}_{\alpha;a}(p,q)$ and $\lambda_{a}^{\alpha}(p)=\tau^{\alpha}_a \frac{p_x}{m}$. 
In the following we will use the definition:
\begin{equation}
    R_a(p;q)=G_a(p+\frac{q}{2})G_a(p-\frac{q}{2})
\end{equation}
We have another relation for $\Lambda$:
\begin{equation}
    \Lambda^{\alpha}_a(p,q)=\lambda^{\alpha}_a(p)+\sum_{b,p'} \Gamma_{ab}(p,p';q)R_b(p';q)\lambda^\alpha_b(p')
\end{equation}
where $\Gamma$ is a vertex for four-fermion interaction.

Now for a fixed $q$, we will view $\lambda^\alpha,\Lambda^\alpha, R, \Gamma$ as matrices with indices $i,j=(a,p)$.  $\lambda^\alpha, \Lambda^\alpha, R$ are all diagonal matrices.   In this language, we have
\begin{equation}
    \Lambda^\alpha_{aa}=\lambda^\alpha_{aa}+ \sum_b \Gamma_{ab} R_{bb} \lambda_{bb}
    \label{eq:Lambda}
\end{equation}
where $\Lambda^\alpha_{aa}$ is a block-matrix specified by $aa$ component.
$\Gamma$ can be derived from the following self-consistent equation:
\begin{equation}
    \Gamma(q)=\gamma(q)+\gamma(q) R(q) \Gamma(q)
    \label{eq:Gamma}
\end{equation}
where $\gamma(q)=\gamma_{ab}(p,p';q)$ is the two-particle irreducible electron-hole interaction.  The above equation should be understood as a matrix equation.
For each fixed $p,p'$, $\gamma(p,p';q)=\gamma(p,p',-q)$ because every bare interaction has the property $V(\mathbf q)=V(-\mathbf q)$. It is also obvious that $R_a(q)=R_a(-q)$. Then we can prove $\Gamma(q)=\Gamma(-q)$ from (\ref{eq:Gamma}) and $\Lambda^\alpha_{aa}(q)=\Lambda^\alpha_{aa}(-q)$ from (\ref{eq:Lambda}).  We can then use the following expansion around $q=0$:
\begin{align}
    \gamma(q)&=\gamma_0+\gamma_{2} q^2 \notag\\
    \Lambda^{\alpha}&=\Lambda^\alpha_0+\Lambda^\alpha_{2}q^2\notag\\
    R&=R_0+R_{2}q^2 \notag\\
    \Gamma&=\Gamma+\Gamma_2 q^2
\end{align}
We also define $\Pi_\alpha(q)=P_\alpha q^2$, then
\begin{equation}
    P_\alpha=\text{Tr} (\lambda^\alpha R_0 \Lambda^\alpha_2+\lambda^\alpha R_2 \Lambda^\alpha_0)
    \label{eq:P}
\end{equation}
With some algebra, one can reach
\begin{align}
\Gamma_0^{-1}&=\gamma_0^{-1}-R_0\notag\\
\Gamma_2&=\Gamma_0 R_2 \Gamma_0+(1+\Gamma_0 R_0)\gamma_2 (1+\Gamma_0 R_0)\notag\\
\Lambda^\alpha_{0;aa}&=\lambda^\alpha_{aa}+\sum_b \Gamma_{0;ab} R_{0;bb} \lambda_{bb}\notag\\
\Lambda^\alpha_{2;aa} &=\sum_b \Gamma_{0;ab} R_{2;bb} \lambda^\alpha_{bb}+ \sum_b \Gamma_{2;ab} R_{0;bb}\lambda^\alpha_{bb} 
\label{eq:Gamma_Lambda}
\end{align}
Substituting (\ref{eq:Gamma_Lambda}) into (\ref{eq:P}), we get $P_{\alpha}$. To simplify the expression, we assume the following: (1) $G_{+}(p)=G_{-}(p)$;  (2) $\Gamma_{ab}(p,p';q)=\Gamma_{ba}(p,p';q)$.  Both assumptions can be guaranteed if there is a $Z_2$ symmetry acting like $i\tau_x$ or $i\tau_y$, which exchanges the flavor index.   This is true for DQCP2 and DQCP3. Then we obtain
\begin{equation}
    P_\alpha=  \sum_a \Lambda^\alpha_{0;a} R_{2;a} \Lambda^\alpha_{0;a} +\sum_{ab}(R_{0} \Lambda^\alpha_{0})_a \gamma_{2;ab} (\Lambda^\alpha_0 R_0)_b 
\end{equation}
From the Ward identity in (\ref{eq:Ward_Identity}) we get
\begin{equation}
    \Lambda^\alpha_{0;a}= \tau^\alpha_a \frac{\partial G^{-1}_a(p)}{\partial p_x}
\end{equation}
We then find $(\Lambda^\alpha_0 R_0)_a=-\tau^{\alpha}_a \frac{\partial G_a(p)}{\partial p_x}$.
Then the final result for the diagmagnetic susceptibility is
\begin{equation}
    P_\alpha= \sum_a  \sum_p \Lambda^\alpha_{0;a}(p) R_{2;a}(p) \Lambda^\alpha_{0;a}(p) + \sum_{ab}  \tau^{\alpha}_a \tau^{\alpha}_b \sum_{p,p'} \frac{\partial G_a(p)}{\partial p_x}\gamma_{2;ab}(p,p') \frac{\partial G_b(p')}{\partial p'_x}
    \label{eq:general_result}
\end{equation}

We are interested in $P_c-P_s$.  Because of the Ward identity $\Lambda^\alpha_{0;a}= \tau^\alpha_a \frac{\partial G^{-1}_a(p)}{\partial p_x}$, the first term is the same for $\alpha=c$ and $\alpha_s$ given $(\tau^{\alpha}_a)^2=1$.  We there obtain the main result
\begin{equation}
    P_c-P_s=4  \sum_{p,p'} \frac{\partial G_+(p)}{\partial p_x}\gamma_{2;+-}(p,p') \frac{\partial G_-(p')}{\partial p'_x}
\end{equation}
for the differences in the diamagnetic susceptibilities of the two gauge fields.

\subsection{Vanishing of $P_c-P_s$ up to three-loop}

\begin{figure}[ht]
    \centering
    \includegraphics[width=0.9\textwidth]{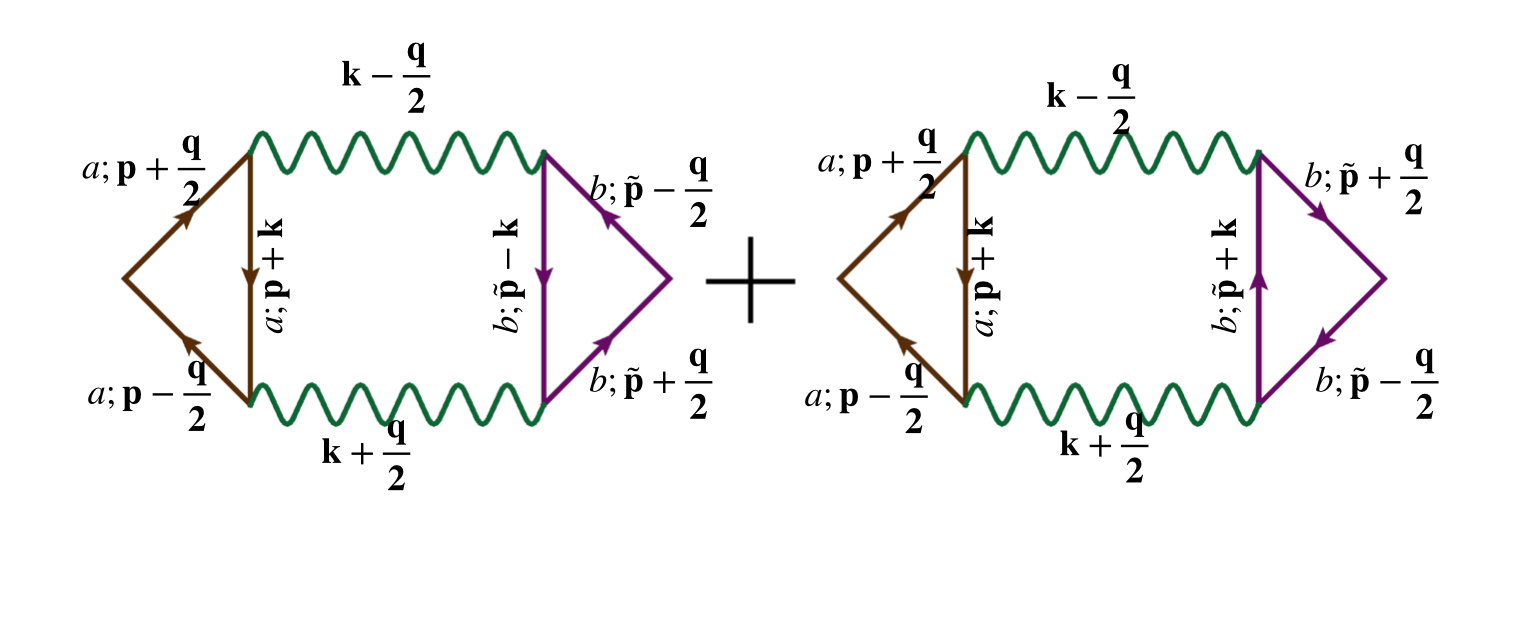}
    \caption{The Aslamazov-Larkin diagram to $\Pi_{ab}(\mathbf q)$, $a,b=+,-$.}
    \label{fig:AL_diagram}
\end{figure}

Recall that $\gamma$ is the electron-hole two-particle irreducible interaction.  The simplest term is just the bare interaction, but it does not provide a non-zero $\gamma_{2}$.   The next order is the Aslamazov-Larkin diagram  at order $V^2$, as shown in Fig.~\ref{fig:AL_diagram}. However, we can prove that $P_c-P_s=0$ up to this order.  The diagram is 
\begin{equation}
\sum_{k} \frac{\partial^2}{\partial q^2} [V(k+\frac{q}{2}) V(k-\frac{q}{2})]  \sum_p \frac{\partial G(p)}{\partial p_x} G(p+k) \sum_{\tilde p}\frac{\partial G(\tilde p)}{\partial \tilde p_x} (G(\tilde p+k)+G(\tilde p-k))
\end{equation}
We have 
\begin{align}
    &\sum_{\tilde p} \frac{\partial G(\tilde p)}{\partial \tilde p_x} G(\tilde p+k)= -\partial_k \sum_{\tilde p} G(\tilde p)G(\tilde p+k)=-\partial_k F(k)\notag\\
    &\sum_{\tilde p} \frac{\partial G(\tilde p)}{\partial \tilde p_x} G(\tilde p+k)= \partial_k \sum_{\tilde p} G(\tilde p)G(\tilde p-k)=\partial_k F(-k)
\end{align}
where $F(k)=\sum_{\tilde p} G(\tilde p)G(\tilde p+k)$.  We can prove $F(k)=F(-k)$, where $-k=(k_0,-\vec{k})$.  This is because $G(\tilde p)=G(-\tilde p)$ as long as we have $\xi(k)=\xi(-k)$.
Then the contribution  vanishes.  We conclude that $P_c-P_s=0$ up to $O(V^2)$.  

Actually, one can finds that there is even no correction to $P_c$ or $P_s$ individualy.  This is because the first term in (\ref{eq:general_result}) gives the following term at order $O(V^2)$:
\begin{align}
    &\sum_p \lambda(p) R_0(p) \gamma_0(p,\tilde p) R_2(\tilde p) \lambda(\tilde p)\notag\\
    &=\sum_{k} V(k) V(k)  \sum_p -\frac{\partial G_0(p)}{\partial p_x} (G_0(p+k)+G_0(p-k))  \sum_{\tilde p} (G(\tilde p+k)+G(\tilde p-k)) R_2(\tilde p)\lambda(p')
\end{align}
Again the summation of $p$ gives $\partial_k F(k)-\partial_k F(-k)=0$.

In summary, there is no correction to the result $\Pi_c(q)=\Pi_s(q)=\chi_f |q|^2$ up to $O(V^2)$.

\section{RG flow of the ghost Fermi surface coupled to a $U(1)\times U(1)$ gauge field \label{append:RG_flow}}

At the QCP, the Higgs boson decouples, and we only need to consider the coupling between the ghost Fermi surface and the $U(1)_S\times U(1)_1$ gauge field. We consider the action
\begin{equation}
  S=S_a+S_\Psi
\end{equation}
with
\begin{equation}
   S_a= \frac{1}{2} \int \frac{d \omega d^2 q}{(2\pi)^3 } \left[ \left( \frac{1}{e_c^2}|q_y|^{1+\epsilon} +\kappa_0 \frac{|\omega|}{|q_y|} \right) |a_1(\omega,\mathbf q)|^2+ \left(\frac{1}{e_s^2}|q_y|^{1+\epsilon}+\kappa_0 \frac{|\omega|}{|q_y|} \right) |a_S(\omega,\mathbf q)|^2 \right]
\end{equation}
and
\bea
  S_\psi &=& \int \frac{d\omega d^2 k}{(2\pi)^3}  \Psi^\dagger( i\omega-\upsilon_F k_x-\kappa k_y^2)\Psi
  \nonumber \\
  &~&~~~+  \upsilon_F \int \frac{d^3 q}{(2\pi)^3}  \int \frac{d\omega d^2 k}{(2\pi)^3}  \Psi^\dagger(\mathbf k+\mathbf q)(a_1 (\mathbf q) + a_S (\mathbf q) \tau^z ) \Psi(\mathbf k)
\eea
At the QCP, we have $\epsilon=0$.  When the Higgs boson mass is large, we have $\epsilon=1$ We have scaling $[k_y]=\frac{1}{2}$, $[k_x]=1$, $[\omega]=1$,$[\Psi]=-\frac{7}{4}$, $[a]=[\alpha]=-\frac{3}{2}$, $[e_c^2]=[e_s^2]=\frac{\epsilon}{2}$, $[\upsilon_F]=0$.  Actually the only meaningful coupling is $\alpha_c=\frac{e_c^2 \upsilon_F}{4\pi^2}$ and $\alpha_s=\frac{e_s^2 \upsilon_F}{4\pi^2}$.  From the naive scaling we get $[\alpha_c]=[\alpha_s]=\frac{\epsilon}{2}$.   At $\epsilon=0$ the coupling is marginal and there is hope to do controlled calculation.

Next we perform a renormalization group analysis using $\epsilon$ expansion. It is useful to introduce the redefinition:
\begin{align}
\Psi_0&=Z^{1/2}\Psi \notag\\
\upsilon^0_F&=Z_{\upsilon_F}\upsilon_F \notag\\
e^0_c &=\mu^{\frac{\epsilon}{4}} e_c Z_{e_c}\notag\\
e^0_s &=\mu^{\frac{\epsilon}{4}} e_s Z_{e_s}\notag\\
a_1^0&=Z_{a_1}a_1 \notag\\
a_S^0&=Z_{a_S} a_S
\end{align}
and then we can rewrite the original action as
\begin{align}
S &= \frac{1}{2} \int \frac{d \omega d^2 q}{(2\pi)^3 }  \Biggl[ \left(\frac{Z_{a_1}^2}{ \mu^{\frac{\epsilon}{2}} Z_{e_c}^2 e_c^2}|q_y|^{1+\epsilon} + Z_{a_1}^2\kappa_0 \frac{|\omega|}{|q_y|} \right) |a_1(\omega,\mathbf q)|^2 \notag \\
& +\sum_{a=1,2,3} \left(\frac{Z_{a_S}^2}{\mu^{\frac{\epsilon}{2}} Z_{e_s}^2e_s^2}|q_y|^{1+\epsilon}+Z_{a_S}^2\kappa_0 \frac{|\omega|}{|q_y|} \right) |\alpha_s^a(\omega,\mathbf q)|^2 \Biggr] \notag\\
&+\int \frac{d\omega d^2 k }{(2\pi)^3} \Psi^\dagger( iZ\omega-Z Z_{\upsilon_F}\upsilon_F k_x- Z Z_{\upsilon_F}\kappa k_y^2)\Psi\notag\\
&+\upsilon_F Z Z_{\upsilon_F} \int \frac{d^3 q}{(2\pi)^3}  \int \frac{d\omega d^2 k}{(2\pi)^3}  \Psi^\dagger(\mathbf k+\mathbf q)(Z_{a_1} a_1 (\mathbf q)+ Z_{a_S} a_S (\mathbf q) \tau^z ) \Psi(\mathbf k)
\end{align}
From the Ward identity, we expect $Z_{a_1}=1$ and $Z_{a_S}=1$.  Hence the fermion-gauge field vertex correction should be purely from $Z Z_{\upsilon_F}$. As we will see, $Z Z_{\upsilon_F}=1$, implying that there is no vertex correction.  When $\epsilon<1$, we expect $Z_{e_c}=Z_{e_s}=1$ because the non-analytic form $|q_y|^{1+\epsilon}$ can not be renormalized.  Therefore the only important renormalization is from $Z=Z^{-1}_{\upsilon_F}$.

The fermion self-energy at one-loop order is
\begin{align}
\Sigma(i\omega)&=- \frac{e_c^2 \upsilon_F^2}{(2\pi)^3}\int d q_0 d^2 q  \frac{1}{|q_y|^{1+\epsilon}+\kappa_0 e_c^2 \frac{|q_0|}{|q_y|}} \frac{1}{i\omega+iq_0-\upsilon_F (k_x+q_x)+\kappa (k_y+q_y)^2}\notag\\
&+\frac{e_s^2 \upsilon_F^2}{(2\pi)^3}\int d q_0 d^2 q  \frac{1}{|q_y|^{1+\epsilon}+\kappa_0 e_s^2 \frac{|q_0|}{|q_y|}} \frac{1}{i\omega+iq_0-\upsilon_F (k_x+q_x)+\kappa (k_y+q_y)^2} \notag\\
&= \frac{\alpha_c}{2} \int dq_0 dq_y \frac{i\sign(\omega+q_0)}{|q_y|^{1+\epsilon}+\kappa_0 e_c^2 \frac{|q_0|}{|q_y|}}+\frac{\alpha_s}{2} \int dq_0 dq_y\frac{i\sign(\omega+q_0)}{|q_y|^{1+\epsilon}+\kappa_0 e_s^2 \frac{|q_0|}{|q_y|}}\notag\\
&=(\alpha_c+\alpha_s) \frac{1}{\epsilon} \int dq_0  i\sign(\omega+q_0)  +... \notag\\
&=2(\alpha_c+\alpha_s) i\omega \frac{1}{\epsilon} \label{selfu1}
\end{align}
In the above we only keep the divergent part $O({1}/{\epsilon})$. To cancel the divergent part, we need 
\begin{align}
Z=Z^{-1}_{\upsilon_F}=1-2(\alpha_c+\alpha_s)\frac{1}{\epsilon} \label{Zu1}
\end{align}

Next, we show explicitly that the vertex correction vanishes \cite{Mross:2010rd}. For simplicity we use $a_1$ as an illustration. We have
\begin{align}
\delta \Gamma^c (p_0,p_x,p_y)&=\int \frac{dq^3}{2\pi} \frac{1}{iq_0-\upsilon_F q_x-\kappa_0 q_y^2} \frac{1}{iq_0+i p_0-\upsilon_F(q_x+p_x)-\kappa_0(q_y+p_y)^2} \times\notag\\
&\ \ \ ~~~~~~~~~~~~\left(\frac{\alpha_c\upsilon_F}{|q_y|^{1+\epsilon}+\kappa_0 e_c^2 \frac{|q_0|}{|q_y|}}+\frac{\alpha_s\upsilon_F}{|q_y|^{1+\epsilon}+\kappa_0 e_s^2 \frac{|q_0|}{|q_y|}} \right)\notag\\
&= i \sign(p_0) \int dq_y \int_0^{|p_0|} d q_0  \left(\frac{\alpha_c}{|q_y|^{1+\epsilon}+\kappa_0 e_c^2 \frac{|q_0|}{|q_y|}}+\frac{\alpha_s}{|q_y|^{1+\epsilon}+\kappa_0 e_s^2 \frac{|q_0|}{|q_y|}} \right) \times \notag \\
&\ \ \ ~~~~~~~~~~~~\frac{1}{ip_0-\upsilon_F p_x-2 \kappa_0 p_y q_y-\kappa_0 p_y^2} 
\end{align}
where $(p_0,p_x,p_y)$ is the external momentum of the photon at the vertex.  We assume that the $(\omega,k_x,k_y)=(0,0,0)$ for one external fermion.   In the first step we integrate $q_x$ and get a factor $\sign(p_0+q_0)-\sign(q_0)$, which is equal to $2$ for $q_0\in [-p_0,0]$ and zero elsewhere.
It is easy to find that $\delta \Gamma^c (p_0,p_x,p_y)=0$ in the $p_0=0$, but $p_x,p_y$ finite limit.   Thus we conclude that there is no vertex correction.  The same conclusion holds for the vertex corresponding to $a_S$.

Finally, we can get the beta function $\beta(\alpha_c)=-{d\alpha_c}/{d\log\mu}$ and $\beta(\alpha_s)=-{d\alpha_s}/{d\log \mu}$ (note, this is the negative of the usual definition) from the relation $ \alpha_c^0=\mu^{\frac{\epsilon}{2}} \alpha_c Z_{e_c}^2 Z_{\upsilon_F}$ and  $\alpha_s^0=\mu^{\frac{\epsilon}{2}} \alpha_s Z_{e_s}^2 Z_{\upsilon_F}$.
We have equations:
\begin{align}
0&=-\frac{\epsilon}{2}+\frac{1}{\alpha_c}\beta(\alpha_c)+\frac{2}{Z_{e_c}} \left(\frac{\partial Z_{e_c}}{\partial \alpha_c}\beta(\alpha_c)+\frac{\partial Z_{e_c}}{\partial \alpha_s}\beta(\alpha_s)\right)+\frac{1}{Z_{\upsilon_F}} \left(\frac{\partial Z_{\upsilon_F}}{\partial \alpha_c}\beta(\alpha_c)+\frac{\partial Z_{\upsilon_F}}{\partial \alpha_s}\beta(\alpha_s)\right)\notag\\
0&=-\frac{\epsilon}{2}+\frac{1}{\alpha_s}\beta(\alpha_s)+\frac{2}{Z_{e_s}} \left(\frac{\partial Z_{e_s}}{\partial \alpha_c}\beta(\alpha_c)+\frac{\partial Z_{e_s}}{\partial \alpha_s}\beta(\alpha_s)\right)+\frac{1}{Z_{\upsilon_F}} \left(\frac{\partial Z_{\upsilon_F}}{\partial \alpha_c}\beta(\alpha_c)+\frac{\partial Z_{\upsilon_F}}{\partial \alpha_s}\beta(\alpha_s)\right)
\end{align}
The above equations can be written as:
\begin{align}
\left[\frac{1}{\alpha_c}+\frac{2}{\epsilon+2(\alpha_c+\alpha_s)}\right]\beta(\alpha_c)+\frac{2}{\epsilon+2(\alpha_c+\alpha_s)}\beta(\alpha_s)&=\frac{\epsilon}{2}\notag\\
\frac{2}{\epsilon+2(\alpha_c+\alpha_s)} \beta(\alpha_c)+\left[\frac{1}{\alpha_s}+\frac{2}{\epsilon+2(\alpha_c+\alpha_s)}\right]\beta(\alpha_s)&=\frac{\epsilon}{2}\notag\\
\end{align}
The solution is
\begin{align}
\beta(\alpha_c)&=\frac{\epsilon}{2} \alpha_c-\alpha_c(\alpha_c+\alpha_s)\notag\\
\beta(\alpha_s)&=\frac{\epsilon}{2}\alpha_s-\alpha_s(\alpha_c+\alpha_s) \label{betau1}
\end{align}
At the QCP, we expect $\alpha_c(\ell=0)=\alpha_s(\ell=0)={1}/{\sigma_b}$.  Here $\ell=-\log \mu$ is the time of the RG flow. Then according to the above RG flow, $\alpha_c=\alpha_s$ will remain true for any $\ell$.  In the $\ell \rightarrow \infty$ limit we reach the fixed point $\alpha_c=\alpha_s=0$.  As usual the fermion self energy has a $\log$ correction $i\omega \log \omega$ and the specific heat $C/T$ from the ghost fermion also has a $\log T$ correction.

\section{Pairing instability for the FL*-FL transition with $U(2)$ gauge theory}
\label{append:RG_U2}

We can easily generalize the calculation in Appendix~\ref{append:RG_flow} for the $U(1)_S \times U(1)_1$ gauge theory of DQCP2 and DQCP3 to the $SU(2)_S \times U(1)_1$ theory for DQCP1 of the FL*-FL transition. The main change is that the $a_S$ gauge field now has 3 components $a_S^\alpha$, $\alpha=x,y,z$. This has the consequence that the fermion self energy in (\ref{selfu1}) is modified to 
\begin{align}
\Sigma(i\omega)&=- \frac{e_c^2 \upsilon_F^2}{(2\pi)^3}\int d q_0 d^2 q  \frac{1}{|q_y|^{1+\epsilon}+\kappa_0 e_c^2 \frac{|q_0|}{|q_y|}} \frac{1}{i\omega+iq_0-\upsilon_F (k_x+q_x)+\kappa (k_y+q_y)^2}\notag\\
&+\frac{3e_s^2 \upsilon_F^2}{(2\pi)^3}\int d q_0 d^2 q  \frac{1}{|q_y|^{1+\epsilon}+\kappa_0 e_s^2 \frac{|q_0|}{|q_y|}} \frac{1}{i\omega+iq_0-\upsilon_F (k_x+q_x)+\kappa (k_y+q_y)^2} \notag\\
&= \frac{\alpha_c}{2} \int dq_0 dq_y \frac{i\sign(\omega+q_0)}{|q_y|^{1+\epsilon}+\kappa_0 e_c^2 \frac{|q_0|}{|q_y|}}+\frac{3 \alpha_s}{2} \int dq_0 dq_y\frac{i\sign(\omega+q_0)}{|q_y|^{1+\epsilon}+\kappa_0 e_s^2 \frac{|q_0|}{|q_y|}}\notag\\
&=(\alpha_c+3\alpha_s) \frac{1}{\epsilon} \int dq_0  i\sign(\omega+q_0)  +... \notag\\
&=2(\alpha_c+3\alpha_s) i\omega \frac{1}{\epsilon} \label{selfsu2}
\end{align}
In the above, an additional factor of $3$ before $\alpha_s$ is needed because we sum over the three components $a_S^\alpha$. In the fourth line we only keep the divergent part $O(\frac{1}{\epsilon})$.
So the renormalization factors in (\ref{Zu1}) are replaced by
\begin{align}
Z=Z^{-1}_{\upsilon_F}=1-2(\alpha_c+3\alpha_s)\frac{1}{\epsilon} \,. \label{Zsu2}
\end{align}
Following the procedure in Appendix~\ref{append:RG_flow}, we now obtain the $\beta$ functions 
replacing (\ref{betau1})
\begin{align}
\beta(\alpha_c)&=\frac{\epsilon}{2} \alpha_c-\alpha_c(\alpha_c+3\alpha_s)\notag\\
\beta(\alpha_s)&=\frac{\epsilon}{2}\alpha_s-\alpha_s(\alpha_c+3\alpha_s)\,. \label{betasu2}
\end{align}
At the QCP, at one-loop order in the Higgs boson fluctuations $\alpha_c(\ell=0)=\alpha_s(\ell=0)={1}/{\sigma_b}$. However, we don't expect this equality to be obeyed at higher loops, because the $SU(2)_S$ and $U(1)_1$ sectors will behave differently. Nevertheless, it is easy to check from (\ref{betasu2}) that $\alpha_c/\alpha_s = r$ does not flow with $\ell$.

Then according to the above RG flow, $\alpha_c=\alpha_s$ will remain true for any $\ell$. 

\subsection{Pairing instability}

The leading contribution to interaction in BCS channel for ghost Fermi surface $\Psi$ is from exchange of one photon or gluon.  Generically, it is in the form:
\begin{equation}
  S_{BCS}=\int d^2k_i d \omega_i \psi^\dagger_a(k_1)\psi^\dagger_b(-k_1)\psi_d(-k_2)\psi_c(k_2)\big[ V_a(\delta_{a c} \delta_{b d}+\delta_{a d}\delta_{b c})+V_s(\delta_{a c}\delta_{b d}-\delta_{a d}\delta_{b c}) \big] F(k_1-k_2)
\end{equation}
Here $V_a$ is the pairing with odd angular momentum and $V_s$ is the pairing with even angular momentum.  $F(q=k_1-k_2)$ is coming from the integration of the propagator of the photon or gluon.

Next, we decide the contribution from the $U(1)$ gauge field $a_1$ and $SU(2)$ gauge field $a_S$.  The contribution is proportional to  $\sum_\alpha t^\alpha_{a c}t^\alpha_{b d}$, where $t^\alpha$ are the corresponding generators.  For the $U(1)$ gauge field, we just have $t=I$ and $\sum_\alpha t^\alpha_{a c}t^\alpha_{b d}=\delta_{a c} \delta_{b d}$.  Thus we can get $V_a=V_s=\frac{1}{2}$.   For the $SU(2)$ gauge field, one finds that $V_a=\frac{1}{2}$ and $V_s=-\frac{3}{2}$.  If we sum up the contributions from $a_1$ and $a_S$, $V_a$ is always positive, but $V_s$ can be negative.

So we obtain the following flow equation in the even angular momentum sector
\begin{equation}
  \frac{dV_s}{d \ell}=\frac{1}{2}\alpha_c-\frac{3}{2}\alpha_s-V_s^2
\end{equation}
Using the $\ell$ independence of $r=\alpha_c(\ell)/\alpha_s(\ell)$, at the quantum-critical point ($\epsilon=0$) the RG flow equations become
\begin{align}
\frac{dV_s}{d\ell}&=-\frac{(3-r)}{2} \alpha_s -V_s^2\notag\\
\frac{d\alpha_s}{d\ell}&=\frac{\epsilon}{2}\alpha_s- (3+r) \alpha_s^2
\end{align}
There is a fixed point $(\alpha_s^*,\alpha_c^*)=(\frac{\epsilon}{2(3+r)},\frac{r}{2(3+r)}\epsilon)$. The stability of this fixed point now depends upon whether $3-r$ is negative or positive \cite{MMSS14,ZouDeb1}.  When $r<3$, $V_s$ will flow to $-\infty$, leading to pairing of $\Psi$.

The value of $r$ depends upon the ratio of the $SU(2)_S$ and $U(1)_1$ conductivities of $\Psi$ and $\Phi$. At one loop, the conductivities are equal, and so $r=1$. The value $r$ will be modified by vertex correction at higher order, but it may be  unlikely that $r>3$ (in which case there is a regime where the fixed point, and the critical Fermi surface, is stable).   In conclusion we conjecture that the DQCP1 is unstable to pairing and the pairing scale is $\Delta \sim \Lambda e^{-\frac{\pi}{\sqrt{\frac{3-r}{2} \alpha_s^*}}}$.

\section{Marginal Fermi liquid behavior in three dimensions}
\label{app:marginal}

In this Appendix, we describe details of the evaluation of the self energy of the $C$ fermions discussed in Section~\ref{sec:physical}.

The self energy for the boson $\Phi$ is
\begin{align}
  \tilde \Sigma_\Phi(iq_0,\mathbf q)&=g^2 \sum_p G_c(p+\frac{q}{2}) G_\psi(p-\frac{q}{2}) \notag\\
  &= g^2 \sum_p \frac{1}{-ip_0 -i \frac{q_0}{2} +\upsilon_c (p+\frac{q}{2} \cos \theta)}\frac{1}{-i p_0+i\frac{q_0}{2}-\upsilon_\psi(\delta+p-\frac{q}{2} \cos \theta)}\notag\\
  &= g^2 \sum_{\mathbf p} \frac{\Theta[-(p+\frac{q}{2}\cos \theta)]-\Theta[p+\delta-\frac{q}{2} \cos \theta]}{-i q_0+\upsilon_\psi \delta+(\upsilon_c+\upsilon_\psi)p+(\upsilon_c-\upsilon_\psi)\frac{q}{2} \cos \theta}
\end{align}
First, we calculate $\Sigma_\Phi(0,0)$,
\begin{align}
\tilde \Sigma_\Phi(0,0)&=g^2 k_F^2\frac{1}{(2\pi)^3}\int dp d\theta d \varphi \sin \theta  \frac{\Theta(-p)-\Theta(p+\delta)}{\upsilon_\psi \delta+(\upsilon_c+\upsilon_\psi)p} \notag\\
&=\frac{g^2k_F^2}{2\pi^2(\upsilon_c+\upsilon_\psi)}  \log   \frac{\upsilon_c \upsilon_\psi \delta^2}{(\upsilon_c+\upsilon_\psi)^2 \Lambda^2-\upsilon_\psi^2 \delta^2}
\end{align}
This non-zero $\tilde \Sigma_\Phi(0,0)$ can be absorbed in to the chemical potential $\mu_\Phi$ and we only need to focus on $\Sigma_\Phi(iq_0,\mathbf q)=\tilde \Sigma_\Phi(iq_0,\mathbf q)-\tilde \Sigma_\Phi(0,0)$.
\begin{align}
\Sigma_\Phi(iq_0,\mathbf q)=\frac{g^2 k_F^2}{4\pi^2(\upsilon_c+\upsilon_\psi)} \int_0^{\pi} d\theta \sin \theta \log  \frac{|q_0|^2+i(\upsilon_\psi-\upsilon_c)q_0(\delta-q\cos\theta)+\upsilon_c \upsilon_\psi (\delta-q\cos \theta)^2 }{\upsilon_c \upsilon_\psi \delta^2}
\label{eq:integral_boson_self_energy}
\end{align}
In the region $|q_0| \ll \sqrt{\upsilon_c \upsilon_\psi} \delta$ and $q \ll \delta$, we obtain:
\begin{equation}
  \Sigma_\Phi(iq_0,\mathbf q) \approx \frac{g^2k_F^2}{2\pi^2(\upsilon_c+\upsilon_\psi)} \left[ -i \left(\frac{\upsilon_c-\upsilon_\psi}{\upsilon_c} \right) \frac{q_0}{\upsilon_\psi \delta}- \frac{1}{3} \frac{q^2}{\delta^2} \right]
\end{equation}
where we keep only $O(q^2)$ and $O(q_0)$ terms. 
We have $G^{-1}_{\Phi}(0,\mathbf q)=\frac{1}{2m_\Phi}(1-\frac{2m_\Phi}{3 \delta^2}\alpha)q^2$ where $\alpha= \frac{g^2 k_F^2}{2\pi^2(\upsilon_c+\upsilon_\psi)}$.  If $\frac{2m_\Phi}{3 \delta^2} \alpha>1$, then the $\Phi$ will have minimum at non-zero $\mathbf q$.   Let us focus on the $\frac{2m_\Phi}{3 \delta^2} \alpha<1$ case here. 

For the purpose of the next subsection, we want to obtain the imaginary part of $\Sigma_\Phi(q_0,\mathbf q)$ after analytical continuation $iq_0 \rightarrow q_0$.  We assume  $q_0<<\upsilon q$ and $q>>\delta$.   First, let us assume $\upsilon_c=\upsilon_\psi=\upsilon$.   Then  the imaginary part can be obtained by numerical evaluation of the integral in Eq.~\ref{eq:integral_boson_self_energy}. We find 
\begin{equation}
  \Sigma''_\Phi(q_0,\mathbf q)\approx \frac{g^2k_F^2}{4\pi^2 \upsilon} \frac{\pi |q_0|}{|\upsilon q|}
\end{equation}
The result is different for the case with $\upsilon_c>\upsilon_\psi$. For example, we take $\upsilon_\psi=\upsilon$ and $\upsilon_c=2\upsilon$, then
\begin{equation}
  \Sigma''_\Phi(q_0,\mathbf q)\approx \frac{g^2k_F^2}{6\pi^2 \upsilon} \frac{\pi \sqrt{\frac{|q_0|}{2 \upsilon \delta}}}{\frac{|q|}{\delta}}
\end{equation}

We consider the weak coupling regime with $\frac{2m_\Phi}{3 \delta^2} \alpha<1$ first, where $\alpha= \frac{g^2 k_F^2}{2\pi^2(\upsilon_c+\upsilon_\psi)}$.
The scattering rate of quasi particle is obtained as imaginary part of the self energy after analytical continuation $i\omega \rightarrow \omega$. We can also calculate the imaginary part $\Sigma''_C(\mathbf k)$ directly from Fermi's golden rule.
Assuming $\upsilon_c=\upsilon_\psi=\upsilon$,
\begin{equation}
  G_\Phi(\omega,\mathbf q)=\frac{1}{-\omega+\frac{q^2}{2m_\Phi}(1-\frac{2m_\Phi}{3 \delta^2} \alpha )+i \alpha \frac{\pi |\omega|}{\upsilon |q|}}
\end{equation}

In the following we consider the scattering rate of the process $C(\mathbf k) \rightarrow \Phi(\mathbf q)+ \Psi(\mathbf k')$.  Because of the rotation invariance, We can focus on  the point $(k_x,k_y,k_z)=(k_x,0,0)$ .  We need $q_x \sim-\delta$ to compensate the momentum mismatch between $C$ and $\Psi$.  We can then make a redefinition $q_x \rightarrow -\delta+q_x$, and then
\begin{equation}
  G_\Phi(\omega,\mathbf q)=\frac{1}{-\omega+\frac{\delta^2+2 q_x \delta+q_y^2+q_z^2}{2m_\Phi}(1-\frac{2m_\Phi}{3 \delta^2} \alpha )+i \alpha \frac{\pi |\omega|}{\upsilon |q|}}
\end{equation}
As the dominant process is from $q_x \sim \frac{q_y^2}{2 m_\psi}$, we can ignore the $q_x$ dependence. We will consider an energy scale larger than $E_0\sim \frac{\delta^2}{2m_\Phi}$ and thus we can focus on the regime $q_y,q_z>>\delta$ and set $\delta=0$.  We can also ignore the $-\omega$ term because the dominant region is from $\frac{\omega}{q} \sim q^2$ and $\omega \sim q^3<<q^2$. Therefore, we can approximate
\begin{equation}
  G_\Phi(\omega,\mathbf q)=\frac{1}{\frac{q_y^2+q_z^2}{2m_\Phi}(1-\frac{2m_\Phi}{3 \delta^2} \alpha )+i \alpha \frac{\pi |\omega|}{\upsilon \sqrt{q_y^2+q_z^2}}}
\end{equation}
In the $\frac{2m_\Phi}{3 \delta^2} \alpha<1$ regime,
\begin{equation}
  \text{Im} G_\Phi(\omega,\mathbf q)=\frac{\pi \alpha \frac{\omega}{\upsilon q}}{\frac{\alpha^2 \pi^2 \omega^2}{\upsilon^2 q^2}+(\frac{q^2}{2\tilde m_\Phi})^2}
\end{equation}
where $\tilde m_\Phi=\frac{m_\Phi}{1-\frac{2m_\Phi}{3 \delta^2} \alpha}$ and $q=\sqrt{q_y^2+q_z^2}$.

Now we can compute the imaginary part of the $C$ electron self energy
\begin{align}
    \Sigma''_C(\mathbf k)&= g^2 \int d\omega \int \frac{d^3 k'}{(2\pi)^3}  \delta(\xi_c(\mathbf k)-\xi_\psi(\mathbf k')-\omega) \text{Im} G_\Phi(\omega,\mathbf q=\mathbf{k'-k})\notag\\
    &~~~(1+n_B(\omega))(1-f(\xi_\psi(k')))\notag\\
    &=\frac{g^2 k_F^2}{8\pi^3 \upsilon_\psi} \int d\omega \int d\theta d\varphi  \int d\xi_\psi  \sin \theta \delta(\xi_c(\mathbf k)-\xi_\psi-\omega) \text{Im} G_\Phi(\omega,q_y=k_F \sin \theta \varphi, q_z=k_F \theta)\notag\\
    &~~~  (1+n_B(\omega))(1-f(\xi_\psi))\notag\\
    &=\frac{g^2 }{8\pi^3 \upsilon_\psi} \int d\omega \int dq_y \int dq_z  \text{Im} G(\omega,\mathbf q) \theta(\omega)\theta(\xi_c(\mathbf k)-\omega)\notag\\
    &=\frac{g^2}{4\pi^2 \upsilon_\psi } \int_0^{\xi_c(\mathbf k)}d\omega  \int_0^{\infty}dq  q \frac{\alpha \frac{\omega}{\upsilon q}}{\frac{\alpha^2 \omega^2}{\upsilon^2 q^2}+(\frac{q^2}{2\tilde m_\Phi})^2}\notag\\
    &=f \xi_c(\mathbf k)
\end{align}
where
\begin{equation}
  f=\frac{\alpha 2\tilde m_\Phi}{k_F^2}  \int_0^{\infty} d\tilde q \frac{\tilde q^2}{1+\tilde q^6}=\frac{\alpha  2\tilde m_\Phi \pi}{6k_F^2}
\end{equation}
We can also use $
  \Sigma''_\Phi(q_0,\mathbf q)\approx \frac{g^2k_F^2}{6\pi^2 \upsilon} \frac{\sqrt{\frac{|q_0|}{2 \upsilon \delta}}}{\frac{|q|}{\delta}}
$ for the case with $\upsilon_c=\upsilon, \upsilon_\psi=2\upsilon$. But a very similar result will be reached in three dimension.
We find that the scattering rate $(\frac{1}{\tau})_{qp}=f \xi_c(\mathbf k)$ when $\chi_c(\mathbf k)>E_0 \sim \frac{\delta^2}{2 m_\Phi}$ and it is natural to expect $(\frac{1}{\tau})_{qp}=f k_B T$ at finite temperature $T>E_0\sim \frac{\delta^{2}}{2m_\Phi}$.  $E_0$ can be quite small even for large $\delta$ if the boson $\Phi$ has a flat band. 
With redefinition $\tilde \alpha=\frac{2m_\Phi}{3 \delta^2} \alpha$, we get
\begin{equation}
    f=\frac{\pi}{2} \frac{\delta^2}{k_F^2} \frac{\tilde \alpha}{1-\tilde \alpha}
\end{equation}
Therefore the coefficient $f$ can be quite large when $\tilde \alpha$ approaches $1$, where the renormalized mass of boson $\tilde m_\Phi$ diverges.

\section{Relation to Hertz-Millis theory}
\label{app:hm}

Here we want to comment on the relation between DQCP2 and the conventional Hertz-Millis theory \cite{hertz,millis}. In principle the two phases separated by these two QCPs are the same. The DQCP2 provides an example of a beyond Landau theory for a Landau allowed phase transition.

In our theory, the DQCP2 is actually a critical line specified by a parameter $M_1$, which controls the size of the ghost Fermi surface. If we increase $M_1>M_{1;c}$, the ghost Fermi surface size becomes zero. In this case, we now show that our DQCP2 will reduce to the conventional Hertz-Millis theory.

When $M_1>M_{1;c}$, $\Psi$ is gapped out by the $M_1$ term. Then $\Phi_+$ and $\Phi_-$ are two  CP$^1$ QED theories, which are equivalent to two 3D $O(3)$ theories if there are no crossing terms between $\Phi_+$ and $\Phi_-$.  Because of the monopole operator, $\Phi$ is confined and the critical theory should be formed by gauge invariant operator.   
By defining $\vec{n}_+=\Phi_+^\dagger \vec \sigma \Phi_+$ and $\vec{n}_-=\Phi_-^\dagger \vec \sigma \Phi_-$, we can use (\ref{eq:Higgs_boson:DQCP2}) to write the Lagrangian as:
\begin{align}
  L_{\mathbf n} &=|\partial_\mu \vec n_c|^2+m_c^2|\vec n_c|^2+|\partial_\mu \vec n_s|^2+m_s^2 |\vec n_s|^2 \notag \\ &+\lambda_c |\vec n_c|^4+\lambda_s |\vec n_s|^4+\lambda_{1} |\vec n_c|^2 |\vec n_s|^2+\lambda_2 |\vec n_c \cdot \vec n_s|^2
\end{align}
where $\vec n_c=\vec n_+ +\vec n_-$ and $\vec n_s=\vec n_+-\vec n_-$. The translation $T_x $ transforms as $\vec n_1 \leftrightarrow \vec n_2$ and thus $\vec n_c \rightarrow \vec n_c$ and $\vec n_s \rightarrow -\vec n_s$. As a result there is no term like $\vec n_c \cdot \vec n_s$.  Physically we can view $\vec n_c$ as FM order parameter and $\vec n_s$ as AF order parameter.

In principle $\vec n_c$ and $\vec n_s$ can be disordered at different couplings. For example, we can have a critical point where $m_s^2$ goes to zero while $m_c^2>0$. This is just the conventional Hertz-Millis theory for an antiferromagnetic critical point.  One can also add the CDW order parameter $\Phi^\dagger \tau_z \Phi$ to the critical theory. The point is that only one order parameter can condense now without fine tuning. Thus the DQCP2 will be reduced to the conventional symmetry breaking transition corresponding to AF, FM or CDW order.

For DQCP3, it is also a critical line specified by $M_1$.  When we tune $M_1$, the ghost Fermi surface is gapped out through a Lifshitz transition. The Lifshitz transition corresponds to a multicritical point (the red point in Fig.~\ref{fig:global_phase_diagram}.  After the transition, the ghost Fermi surface disappears.  Then the monopole of  the gauge field becomes important and effectives gap out the critical boson $\varphi_+,\varphi_-$. Thus there is no phase transition anymore.
\bibliography{kondo}

\end{document}